# The FORA Fog Computing Platform for Industrial IoT*

Paul Pop[a], Bahram Zarrin[a], Mohammadreza Barzegaran[a], Stefan Schulte[b], Sasikumar Punnekkat[c], Jan Ruh[d] and Wilfried Steiner[d]

[a]*DTU Compute, Technical University of Denmark, 2800 Kgs. Lyngby, Denmark*
[b]*Distributed Systems Group, Vienna University of Technology, Karlsplatz 13, 1040 Wien, Austria*
[c]*Dependable Software Engineering, Mälardalen University, Högskoleplan 1, 722 20 Västerås, Sweden*
[d]*TTTech Labs, TTTech Computertechnik AG, Schoenbrunner Strasse 7 1040 Vienna*

ABSTRACT

Industry 4.0 will only become a reality through the convergence of Operational and Information Technologies (OT & IT), which use different computation and communication technologies. Cloud Computing cannot be used for OT involving industrial applications, since it cannot guarantee stringent non-functional requirements, e.g., dependability, trustworthiness and timeliness. Instead, a new computing paradigm, called *Fog Computing*, is envisioned as an architectural means to realize the IT/OT convergence. In this paper we propose a Fog Computing Platform (FCP) reference architecture targeting Industrial IoT applications. The FCP is based on: deterministic virtualization that reduces the effort required for safety and security assurance; middleware for supporting both critical control and dynamic Fog applications; deterministic networking and interoperability, using open standards such as IEEE 802.1 Time-Sensitive Networking (TSN) and OPC Unified Architecture (OPC UA); mechanisms for resource management and orchestration; and services for security, fault tolerance and distributed machine learning. We propose a methodology for the definition and the evaluation of the reference architecture. We use the Architecture Analysis Design Language (AADL) to model the FCP reference architecture, and a set of industrial use cases to evaluate its suitability for the Industrial IoT area.

## 1. Introduction

We are at the beginning of a new industrial revolution (Industry 4.0), which will bring increased productivity and flexibility, mass customization, reduced time-to-market, improved product quality, innovations and new business models. Industrial Internet of Things (IIoT, also called Industrial Internet) is a key enabling technology for Industry 4.0, where the focus is on interconnected machines [54]. IIoT is providing the infrastructure that underpins our Smart Society (Smart Energy Grid, Smart Cities, Smart and Green Mobility, Smart Manufacturing, etc.), providing solutions for several societal challenges.

However, Industry 4.0 will only become a reality through the convergence of Operational and Information Technologies (OT & IT), which are currently separated in a hierarchical pyramid (Purdue Reference Model [95], see Fig. 1) and use different computation and communication technologies. OT consists of cyber-physical systems that monitor and control physical processes that manage, e.g., automated manufacturing, critical infrastructures, smart buildings and smart cities. These application areas are typically safety-critical and real-time, requiring guaranteed extra-functional properties, such as, real-time behavior, reliability, availability, safety, and security and often required to show compliance to industry specific standards. OT uses proprietary solutions implemented with barriers between each level in the pyramid in Fig. 1, imposing severe restrictions on the information flow.

Instead, a new paradigm, called Fog Computing, is envisioned as an architectural means to realize the IT/OT convergence in Industrial IoT [14]. *Fog Computing* is a "system-level architecture that distributes resources and services of computing, storage, control and networking anywhere along the continuum from Cloud to Things" [65]. With Fog

*The research leading to these results has received funding from the European Union's Horizon 2020 research and innovation programme under the Marie Skłodowska-Curie grant agreement No. 764785, FORA—Fog Computing for Robotics and Industrial Automation. We would like to thank the FORA Ph.D. students for their contribution to this work, in alphabetical order, Cosmin Florin Avasalcai, Zeinab Valojerdi Bakhshi, Patrick Heinrich Denzler, Nitin Desai, Eleftherios Kyriakakis, Marine Kadar, Vasileios Karagiannis, Jia Qian, Alexandre Silva Venito, Mohammed Salman Shaik, Václav Struhár, Koen Pieter Tange, Dao Van-Lan, as well as the WP1 leader DTU Assoc. Prof. Martin Schoeberl and DTU Ph.D. student Niklas Reusch.

paupo@dtu.dk (P. Pop); baza@dtu.dk (B. Zarrin); mohba@dtu.dk (M. Barzegaran); s.schulte@infosys.tuwien.ac.at (S. Schulte); sasikumar.punnekkat@mdh.se (S. Punnekkat); jan.ruh@tttech.com (J. Ruh); wilfried.steiner@tttech.com (W. Steiner)

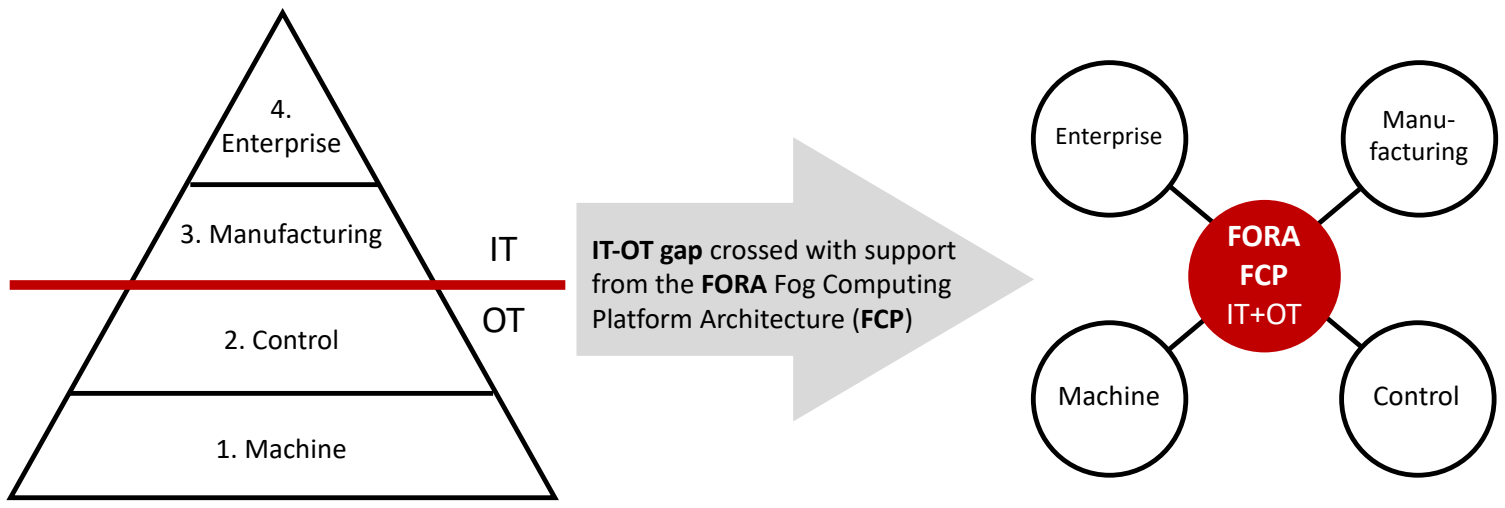

**Figure 1:** Industry 4.0: IT/OT convergence supported by the FORA Fog Computing Platform (FCP)

Computing, communication devices, such as switches and routers are extended with computational and storage resources to enable a variety of communication and computation options (see Fig. 2). Fog Computing will enable a powerful convergence, unification and standardization at the networking, security, data, computing, and control levels. It will lead to improved interoperability, security, more efficient and rich control, and higher manufacturing efficiency and flexibility [15]. Several initiatives are currently working towards realizing this vision [70, 44].

One notable initiative is the European Training Network on Fog Computing for Robotics and Industrial Automation (FORA ETN[1]). The vision is to virtualize the industrial control (which is then implemented as control applications running on a Fog Computing Platform) and achieve the same level of dependability as the one taken for granted in OT. The convergence of IT and OT will be supported by: the increased usage of IP-protocols, e.g., standardized Deterministic Ethernet solutions from IEEE Time-Sensitive Networking (TSN) Task Group [89], upcoming 5G wireless standards [25], and interoperability standards such as OPC Unified Architecture (OPC UA) [58], all integrated into a Fog Computing Platform (FCP), which brings computation, communication and storage closer to the edge of the network. FORA's research objectives focus on: future industrial automation architectures and applications based on Fog Computing, deterministic virtualization and execution, deterministic wired and wireless communication, resource provisioning and resource management, service-oriented architecture solutions, real-time data analytics and security.

### 1.1. Contributions

In this paper we propose a *reference system architecture* for a safe and secure *Fog Computing Platform* (FCP). In our context, an *architecture* captures the fundamental organization of a system (a collection of components performing a specific set of functions) in terms of its components, their relationships to each other, and to the environment, and the principles guiding its design and engineering. A *reference* architecture provides a proven template solution for an architecture for a particular domain; in our case, we are targeting the Industrial IoT area. A Fog Computing *Platform* provides a set of technologies that are used as a base upon which mixed-criticality applications (safety critical, industrial, fog applications) are developed. The proposed reference platform architecture uses open standards and open source, e.g., TSN, OPC UA, 5G and OpenStack for the Edge[2].

We propose a methodology to derive and evaluate the proposed FCP architecture. We use the standardized Architecture Analysis Design Language (AADL) [30] (which is an SAE standard) to model the FCP architecture. We have used several industrial Use Cases (UCs) to evaluate the proposed architecture. We report in the paper in detail the results obtained by using the architecture to implement an industrial conveyor distribution system UC.

The related work is presented in Sect. 2. Sect. 3 introduces the process that has been used to derive the FORA Fog Computing Platform Architecture. The details of the Fog Computing architecture that we propose for Industrial IoT are presented in Sect. 4 using AADL. We discuss the evaluation of the architecture in Sect. 5 where we focus on an industrial use case. Finally, we conclude the paper and provide a discussion in Sect. 6.

## 2. Related Work

Within this section, we discuss the most important related work relevant to the FORA Reference Architecture. Especially, we will have a look at reference architectures in the field of fog computing (Sect. 2.1), the manifold approaches to resource management and optimization that have been presented in recent years (Sect. 2.2), and will discuss the utilization of fog computing in industrial settings (Sect. 2.3).

[1] See the FORA project website for more details, `http://fora-etn.eu`

[2] `http://https://wiki.openstack.org/wiki/Edge_Computing_Group`

## 2.1. Reference Architectures

Especially in the first years of research on fog computing, a number of (reference) architectures have been proposed, starting with the seminal work on fog computing by Bonomi et al. [15, 14]. There, the authors have already presented a layered model, defining fog computing on a high level. The different layers include computational resources in embedded sensors, multi-service edge, the core network and (cloud-based) data centers. An early reference architecture has been presented by Dastjerdi et al. [27], where the authors also divide the fog into a number of hierarchical layers, including IoT devices at the edge of the network, the network itself, cloud services and resources, software-defined resource management, and IoT applications running on top of the fog resources.

A standardized reference architecture for fog computing is the "OpenFog Reference Architecture for Fog Computing" [65] proposed by the OpenFog Consortium, which has later merged with the Industrial Internet Consortium (IIC). This reference architecture has in 2018 been standardized by IEEE as the "1934-2018 IEEE Standard for Adoption of OpenFog Reference Architecture for Fog Computing" [45]. The reference architecture is quite extensive, covering functional pillars of fog computing, a number of use cases for fog computing, as well as the actual reference architecture, provided from different viewpoints. Since the OpenFog Reference Architecture is quite versatile, it does naturally not cover details for industrial settings, which is the focus of the FORA reference architecture. Another important initiative towards standardization in the field of fog and edge computing is led by the European Telecommunications Standards Institute (ETSI), and focusing on Multi-Access Edge Computing (MEC) [37]. Notably, the ETSI activities are very broad and provide a large number of publications which provide a lot of details, up to the level of providing concrete API specifications. In contrast to the work at hand, the ETSI MEC activities rather focus on the edge level, however also taking into account fog aspects. Also, the ETSI MEC activities are use case-agnostic, and therefore not specifically aiming at industry settings, as we take into account.

Apart from the two major reference architectures by the OpenFog Consortium and ETSI, there are a number of more specific reference architectures which are not backed by a large consortium or standardization association: Puliafito et al. [71] present a reference architecture for a "follow-me fog", which follows the user of a mobile device by placing relevant fog services in the proximity of the user, respectively her mobile device. Habibi et al. [42] focus more on the communication aspects by integrating software-defined networking into a fog computing reference architecture. De Brito et al. [16] discuss an extension of the OpenFog Reference Architecture by providing the means for service orchestration.

There are also reference architectures aiming at specific use cases: Mahmud et al. [57] extend basic architecture models for fog computing by adding specific components needed for smart healthcare scenarios, e.g., facilitating interoperability between existing fog clusters. Qi et al. propose a reference architecture for smart manufacturing applications [72]. Here, the authors address an application area comparable to the work at hand. Notably, the reference architecture considers that there is a need for real-time capabilities in fog computing, and that specific resources need to be utilized to achieve this. Their reference architecture contains the means for control, interaction, information integration, and collaboration in smart manufacturing settings. However, in contrast to our work, there is no specific discussion of hard real-time constraints or control applications.

In contrast to the reference architectures discussed so far, our work explicitly aims at industrial settings, taking into account hard real-time constraints and specific needs of control applications.

Apart from the already discussed reference architectures, there have been further discussions on how to structure fog architectures. Mostly, current fog systems make use of a hierarchical architecture, as has already been proposed in the OpenFog Reference Architecture, e.g., [36, 82]. However, there are also architectures which apply a flat structure and apply basic principles from the field of self-organization and peer-to-peer computing, e.g., [18, 75]. While hierarchical approaches are the current state-of-the-art, they possess some potential drawbacks, e.g., that some nodes within a hierarchical fog system may become bottlenecks or single points of failure [51]. Therefore, within the proposed FORA Reference Architecture, we apply a flat, fully distributed architectural style.

## 2.2. Resource Management

Resource management has been a very popular research topic in the field of fog computing in the last few years, with many approaches having been presented so far [11, 43]. The applied methods to manage fog resources are manifold, and range from optimal solutions, e.g., applying mixed-integer linear programming [5], to heuristics, e.g., applying Genetic Algorithms [83]. While most presented solutions are use case-agnostic, there are also specific approaches for vehicular fog computing [99], smart healthcare [41], image processing [97], or industrial settings [92, 94]. According to a systematic literature review by Bellendorf and Mann [11], most approaches to resource management aim at optimizing latency, while energy efficiency also plays a major role. In contrast, cost efficiency, which has been a major research topic in the field of cloud computing [46], is not so prominently discussed in fog computing.

As outlined in the work at hand, hard real-time behavior and safety-related aspects are indispensable in the Industrial IoT. Despite the fact that latency is a major topic in fog resource management, very few studies explicitly aim at hard real-time behavior. For instance, Raagard et al. [69, 74] discuss the runtime reconfiguration of TSN schedules for fog computing, thus supporting applications composed out of hard real-time control tasks. Fizza et al. [33] take into account real-time capabilities during scheduling of tasks and services in the fog. For this, the Earliest Deadline First approach is applied. A similar approach is provided by Gomes et al. [39], aiming at healthcare environments. A conceptual approach to enable real-time fog computing is presented by Kopetz and Poledna [53], where the authors discuss the basic concept of time-triggered VMs. In addition, the support of soft real-time applications is mentioned in quite a large number of studies. For instance, Ning et al. discuss the utilization of fog computing to support real-time traffic management for smart cities [63], and Verma and Sood discuss real-time capabilities in smart healthcare scenarios [91].

### 2.3. Fog Computing in Industrial Settings

We have already discussed some approaches to apply fog computing in industrial settings in the former subsections, focusing on (reference) architectures and resource management for industrial settings, respectively. Within this subsection, we will further discuss related work in this area.

As pointed out above, one important trend in the Industrial IoT is the convergence of OT & IT. Fog computing has been named as an architectural means to achieve this convergence [3, 86]. More concretely, Müller et al. [38, 62] present a reference model for a seamless runtime environment for industrial software, which can thus be deployed in both the fog and the cloud. This model is implemented by orchestrating containers running the single applications. Similarly, Meixner et al. [61] present a framework for automatically placing applications in fog environments, especially aiming at industrial settings.

Apart from this, a number of studies present how to deploy functionalities, which are usually hosted in centralized data centres, more close to the industrial resources, e.g., Cyber-Physical Production Systems (CPPS), by making use of fog resources. Aazam et al. [1] especially discuss the specific requirements of CPPS, and how fog computing architectures can help to overcome their challenges, with a focus on communication, CPS control, (big) data analysis, and sensing. Colelli et al. [22] show how fog resources can be used to improve security in OT and IT networks. Li et al. [56] present an approach to make use of machine learning in the fog in order to improve machine maintenance. Similar work is presented by O'donovan et al. [64]. Fernández-Caramés et al. [32] utilize fog resources in order to enable augmented reality in industrial settings. Zhou et al. [98] discuss how to control CNC machine tools through a fog-based solution. The goal here is once again to achieve (soft) real-time control of CPS in Industrial IoT settings.

While we have only discussed a very limited number of possible applications for fog computing in the Industrial IoT, it can be easily seen that the potential applications cover a very broad spectrum of functionalities. The presented applications could be integrated into the FORA Reference Architecture, thus exploiting the control and real-time capabilities of our work.

## 3. Reference Platform Architecture Definition and Evaluation Methodology

This section introduces briefly the FORA project and proposes a methodology for the reference platform architecture definition and evaluation. FORA is a four-year European Training Network, started in 2017, with the aim of developing a Fog Computing Platform for Industrial IoT. FORA also trains 15 Ph.D. candidates at 7 partner organizations (both academic and industrial), involving also 5 other companies that provide hosting, training and use cases. FORA has a team of over 50 researchers; each Ph.D. candidate has three advisors, both from industry and academia.

Fig. 2 presents a high-level conceptual overview of the FORA Fog Computing Platform architecture. In the left part of the figure, boxes with a red border represent fog nodes, connected with each other and to the Cloud; the thick lines in-between these boxes are the network. Applications (Apps) run in the Fog and Cloud. The FORA project is organized around three main research themes (realized via a corresponding Work Package, WP): Fog Computing Platform (WP1), which virtualizes computation, communication and storage; Resource Management and Middleware (WP2), which uses the platform to provide guarantees for the industrial control applications and novel Fog/Cloud resource management mechanisms (via a Software Manager—SM); and Dependability Services and Application Modeling (WP3), which are vertical services to ensure safety/security aspects and horizontal services to unlock high-value data analytics, implemented as Fog applications. The main building blocks of the platform architecture will be presented in detail in Sect. 4. Here, we discuss the steps of the methodology used for defining the platform architecture, and how the resulted architecture has been evaluated. The details of the evaluation are covered in Sect. 5.

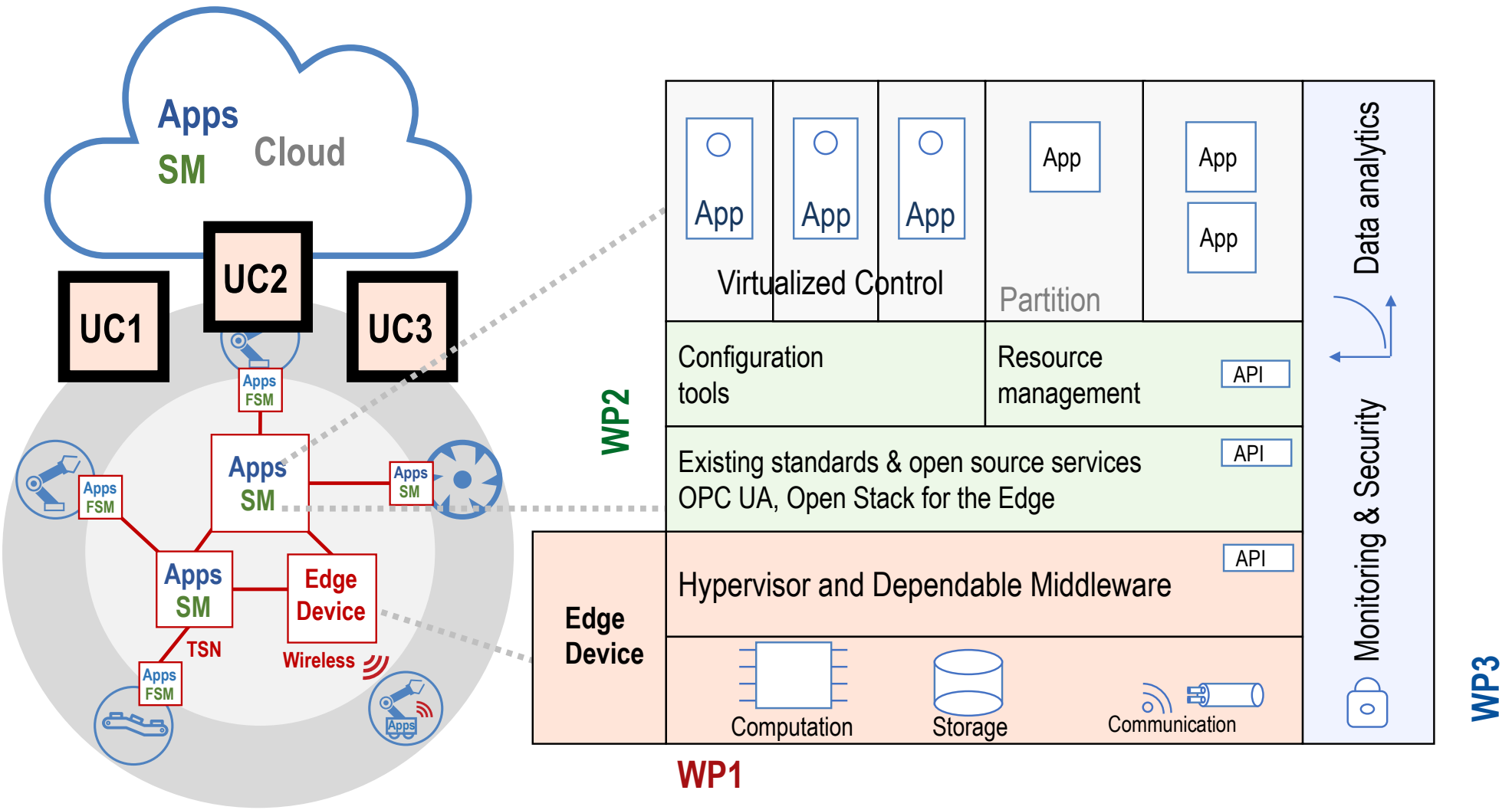

**Figure 2:** FORA Fog Computing Platform concept: UCs and WPs, illustrating the main components

We have identified within FORA several Use Cases (UCs) that are relevant to the FORA organizations and which have a good coverage of the Industrial IoT area: UC1—Electric drives as Fog Nodes in a industrial setting; UC2—Fog-based Industrial Robotic System; UC3—Next generation of machine control using an Edge Platform. A high-level description of these UCs is presented in [34].

The UCs drive the identification of requirements, Key Performance Indicators (KPIs) and UC-specific evaluation metrics. The requirements provide the specific constraints and problems that have to be addressed in the work packages WP1, WP2 and WP3. The evaluation metrics and KPIs establish goals that have to be achieved, and that can be assessed in the evaluation. We have organized the FORA researchers into three teams, one per use case. These use case teams have been asked to identify requirements, KPIs and evaluation metrics. We have provided a template to collect the requirements, consisting of: Requirement ID, Description, Rationale, Abstraction level, AADL component names, and relevance to which WP.

We have collected 10 KPIs and about 80 requirements, see [34] for details. We have consolidated these 80 initial requirements, based on feedback from all the stakeholders, into a coherent set of 47 requirements. The requirements-related documents have been periodically updated during the project based on feedback from the research work and use cases. A partial list of these requirements is presented in Table 1, where we have eliminated the list of the UC-specific and low level requirements due to space limitations.

We have decided to use the Architecture Analysis and Design Language (AADL) [30] to model the FORA FCP architecture, see Sect. 3.1. The FORA AADL models are shared via a "git" repository with all the partners, which contribute to it. Each WP has been tasked with the definition of its respective platform building blocks and the related AADL models, see the sub-sections of Sect. 4.

To evaluate the proposed reference platform architecture, we have: (i) asked the FORA researchers and other stakeholders, to provide feedback on the AADL models that define the reference platform architecture; (ii) used the AADL models to model the three mentioned use cases; and (iii) implemented the use cases as "demonstrator prototypes" to evaluate the ability of our platform architecture to support the design and engineering of IIoT systems. The concrete research outputs of the FORA project, which can be a hardware or software prototype, a method, a tool, a model, etc. are gathered as a set of "Technology Bricks" (TBs). Thus, for (iii), we have used the AADL reference architecture meta-model to model the UCs and integrated these TBs into the demonstrators, one for each UC. We have focused on achieving a high-level of coverage of requirements in the AADL architecture meta-model and a high coverage of AADL components and TBs used in the demonstrators. Sect. 5 presents the evaluation results for the a conveyor

**Table 1**
Selected requirements for the FORA Fog Computing Platform

| ID | Description | Rationale |
|---|---|---|
| R1 | The hardware platform shall provide support for virtualization | Necessary to allow the use and application of modern available hypervisors |
| R2 | The platform shall provide temporal and spatial isolation via hypervisors | Enables hosting of mixed-criticality applications: the lower criticality functions should not impair the safety of the higher criticality functions |
| R3 | The platform shall support lightweight container-based virtualization | This helps to avoid the deployment overhead introduced by the use of hypervisor-based virtualization |
| R4 | The platform shall provide deterministic inter-node communication using standard protocols | To guarantee bounded latency communication between the fog node and its environment, enabling the re-location of critical applications from the machine to the fog node |
| R5 | The platform shall provide reliable and timely wireless communication for the mobile fog nodes | To meet the strict requirements on reliability and latency communication of mobile fog applications such as mobile robots |
| R6 | The platform shall provide fault tolerant communication among the participating compute nodes | In case one or more of the participating compute nodes become unresponsive, the platform should be able to maintain the connectivity of the responsive compute nodes |
| R7 | The platform shall provide standards-based middleware for both industrial and Fog applications | To support the development and deployment of mixed-criticality applications |
| R8 | Each fog node shall broadcast its hardware capabilities | To ensure efficient resource management, any resource manager shall know the available hardware resources within the fog network |
| R9 | The platform shall support the specification and enforcement of security policies | Security policies describe what kind of actions are permissible in a network, and are valuable tools in the prevention and containment of malicious activity |
| R10 | The fog node shall be able to detect errors during its operation and to recover | To ensure the fault-tolerant and high-integrity operation of the safety-related applications |
| R11 | The platform shall be able to run critical control applications | This is needed to virtualize the control equipment (such as PLCs) onto the platform to reduce hardware costs and increase flexibility |
| R12 | The platform shall be compliant with POSIX standards whenever applicable | To ensure portability of applications |
| R13 | Critical real-time tasks should inform their worst-case execution time, period, deadline | To enable the allocation of the necessary resources to the critical tasks to meet their deadlines |
| R14 | The fog nodes shall be able to run data analytics at the edge | To avoid sending all data to the cloud and to support fast optimization and better FCP resource utilization |
| R15 | The platform shall allow for secure retrieval, verification, and execution of software updates | The ability to update has proven itself to be a critical component in the continuous effort to build secure systems |

distribution system demonstrator built for UC1.

## 3.1. Architecture Analysis and Design Language

The Architecture Analysis and Design Language (AADL) is a well-known architecture description language in the domain of real-time embedded systems, which has been introduced by the Society of Automotive Engineers (SAE) [31]. It can model both software and hardware components of a system in a modular and component-oriented approach.

Unlike other modeling languages, e.g., UML and SysML, AADL provides both textual syntax and graphical notations with precise semantics. It introduces a different category of components to model a system, including software components (e.g., data, thread, thread group, subprogram, process), hardware components (e.g., memory, bus, processor, device). It also provides hybrid components (abstract, system) to allow hierarchical system composition and model extension and refinement through the design process.

Similar to object-oriented programming, components can be defined in two levels; component types and component implementations. Component types define the interface of a component, including its external features, e.g., input and output ports. While component implementation specifies the internal elements of a component, such as sub-components and their interactions through connections. Both component types and component implementations can extend other component types or implementations, and each component type can have zero or multiple implemen-tations. Component definitions should be structured within AADL Packages to declare a namespace for them.

Furthermore, AADL allows attaching typed values to the components via properties to specify constraints or characteristics concerning the architecture elements. Several standard properties are available such as timing properties for threads, MIPS capacity for the processors, bandwidth capacity for bus components. Most of AADL analysis tools understand and use these standard properties. It is also possible to define a new set of properties through AADL Property Sets. We use this extensibility mechanism to introduce new properties for specification and configuration of time criticality applications, fog nodes, TSN networks and switches in the FORA reference architecture model.

There are several tools developed for the AADL language to facilitate modeling and analysis of embedded systems from different perspectives such as real-time performance, resource consumption, security, etc. The most well-known one is OSATE [88], which is an open-source Eclipse-based modeling framework. In addition to the modeling environment for the AADL language, it provides a set of plug-ins for validating and analyzing the architecture of the system under study. We have chosen to use AADL as the core language for modeling FORA FCP reference architecture due to its non-ambiguous semantics, human readability, extensibility, and availability of a large set of analysis tools, e.g., scheduler, model checker, flow latency analysis, etc., as OSATE plug-ins.

We have developed the FORA AADLs models based on the standard AADLs components; we have also extended previously proposed meta-models. For example, we have extended the ARINC653 module [93], an AADL annex, which defines virtual processor and virtual buses and targets safety-critical real-time systems, to model hypervisors and address the virtualization and partitioning requirements of the FORA FCP.

As mentioned earlier, FORA has developed a set of Technology Bricks, including a set of methods and tools, e.g., to configure different elements of the FORA FCP platform, such as network topology and routing design [35], streams and task scheduling [10]. We integrate these tools as plug-ins into the OSATE modeling environment in order to facilitate the platform configuration and validation of the FCP use cases. In addition, the models can be analyzed through a set of OSATE-specific plug-ins developed for analyzing safety-critical real-time systems and can be transferred to set up the configuration of the target systems thanks to the Ecore code generation and model transformation mechanisms [85].

Let us illustrate the main concepts of AADL using a simple system consisting of a sensor, an actuator, a computation platform, and an application that consists of two critical control tasks. Fig. 3 presents the architecture of this system as an AADL system component that contains all the system's component instances. This system component represents the top-most level system, which provides the root of the architecture tree, and it must be instantiated to conduct architecture analysis. The sensor and the actuator are modeled as two individual systems with one port to send or receive data. The application is modeled via two processes, which each host one thread; P1 generates data, and P2 consumes the data. Thread components model the active part of an application, and they should be contained in the process components which model the address spaces that contain the threads. We model the platform as an abstract system with two ports for the connectivity purpose to the sensor and actuator, one processor, and a bus to connect them. Abstract components are partially defined components that can be refined during the modeling process. They are especially useful when defining a model as a reference architecture. We choose to use partitioning in order to isolate the execution of the critical tasks on the Processor. Therefore, we model the partitions via two virtual processors introduced by the ARINC653 extension to model a dedicated scheduling domain inside a processor. We use AADL actual processor binding, depicted via blue arrows, to map the processes to the partitions and the partitions to the Processor. For simplicity, we hide the binding visualizations from the AADL diagrams presented in this paper.

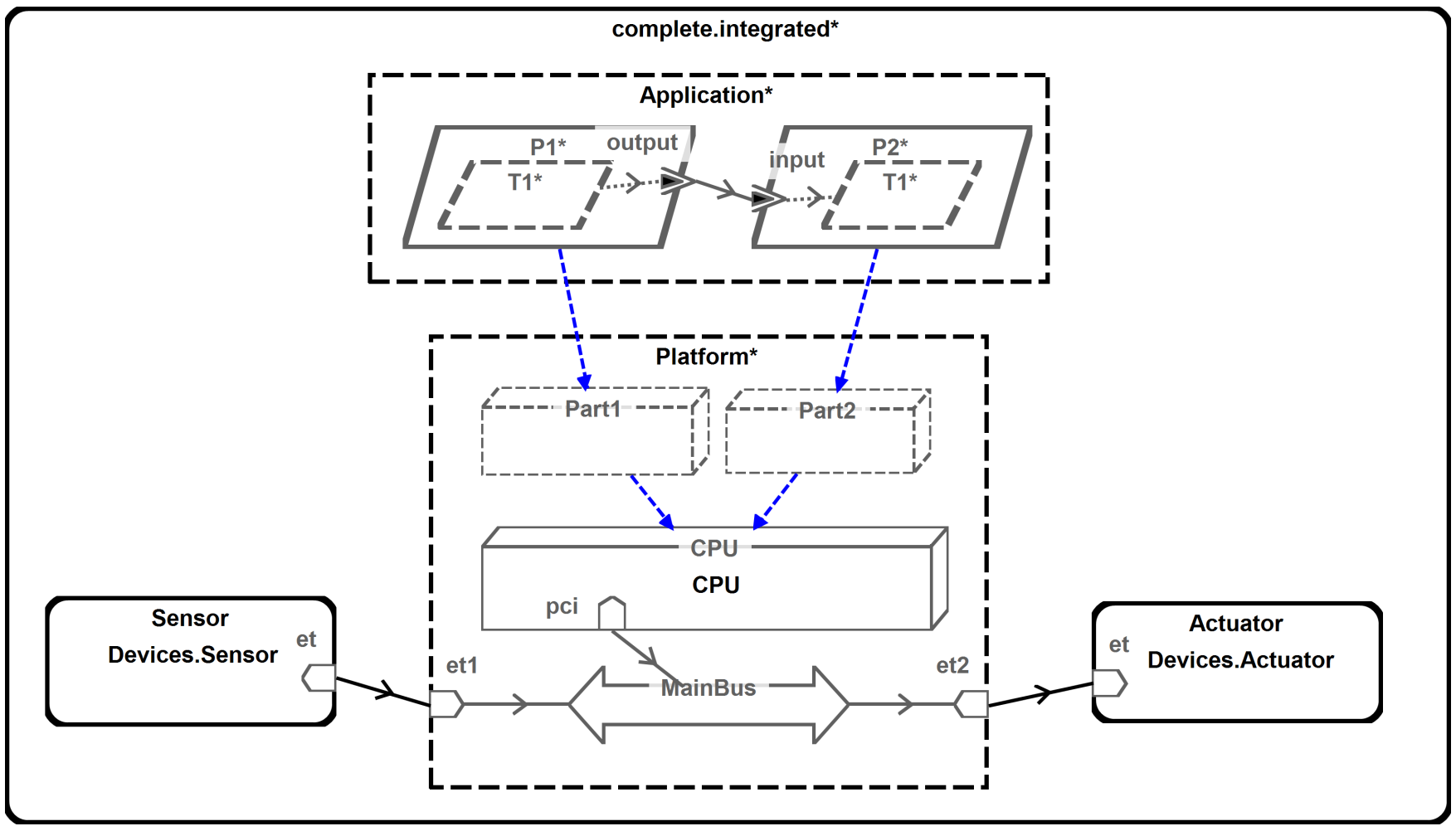

**Figure 3:** An illustration of the main AADL modeling concepts used in this paper

## 4. FORA Fog Computing Platform Reference Architecture Model

The proposed FORA FCP architecture is described in the following three sections from three perspectives: (1) fog computing devices (fog nodes) and the communication among fog nodes, see Sect. 4.1, (2) the mechanisms and techniques for resource management, orchestration and configuration, see Sect. 4.2 and (3) services for dependability, analytics and security, see Sect. 4.3.

An overview of the high-level conceptual architecture is shown in Fig. 4. The foundation of the FCP is the *Fog Node* (FN). In many applications, including industrial automation and robotics, several layers of FNs with differing computation, communication and storage capabilities will evolve, from powerful high-end FNs to low-end FNs with limited resources. Researches have started to propose solutions for the implementation of FNs [15, 70] and fog node solutions have started to be developed by companies [70, 44, 90]. In our case, FN is equipped with a Commercial Of-The-Shelf (COTS) multicore processor (MCP), accelerators, such as FPGAs, for machine learning, and advanced wired and wireless networking capabilities. The FN utilizes its advanced networking capabilities to interact with its environment that includes sensors, actuators, other FNs, and remote Cloud facilities. The initial goal of the IEEE 802.1 TSN Task Group [89] was to provide timing guarantees for demanding applications such as those in the automotive area. Thus, IEEE 802.1 TSN is the ideal technology choice for the fog node's southbound connection. The vision with TSN is to provide a superior technical solution based on open standards. TSN guarantees bounded latency communication between the fog node and its environment. This guarantee enables the re-location of real-time critical tasks from the machine to the fog node. Furthermore, industrial wireless technologies, e.g., WirelessHART or 5G, enable communication with mobile entities or Cloud Facilities in case of remote FN installations, e.g., on offshore oil rigs.

When mixed-criticality functions share the same MCP, they are separated in different virtual machines (partitions) enforced using hardware-supported virtualization [79], based on hypervisors, such as PikeOS [48], ACRN or Xen. The FCP hosts a diverse set of applications with mixed-criticality requirements belonging to both OT and IT domains. The applications are distributed over multiple resources, e.g., low-end FNs integrated in machines, high-end FNs hosting multiple applications of mixed-criticalities, or in the Cloud. This is realised through: A runtime environment and means to orchestrate different applications; cross-layer resource allocation, allocating resources efficiently in volatile scenarios, taking into account different types of resources and a number of non-functional requirements, e.g., latency, cost, security, sensitivity of data; configuration mechanisms and tools, which provision resources such that industrial applications meet the dependability requirements, functioning correctly even in the presence of faults, requiring hence dynamic reconfiguration of computation and communication resources. We build on existing open source software stacks for the edge, e.g., OpenStack for the Edge. The Fog middleware will also build on application layer protocols such as MQTT-SN [84] and CoAP [81] for northbound communication and TSN and OPC UA for southbound communication.

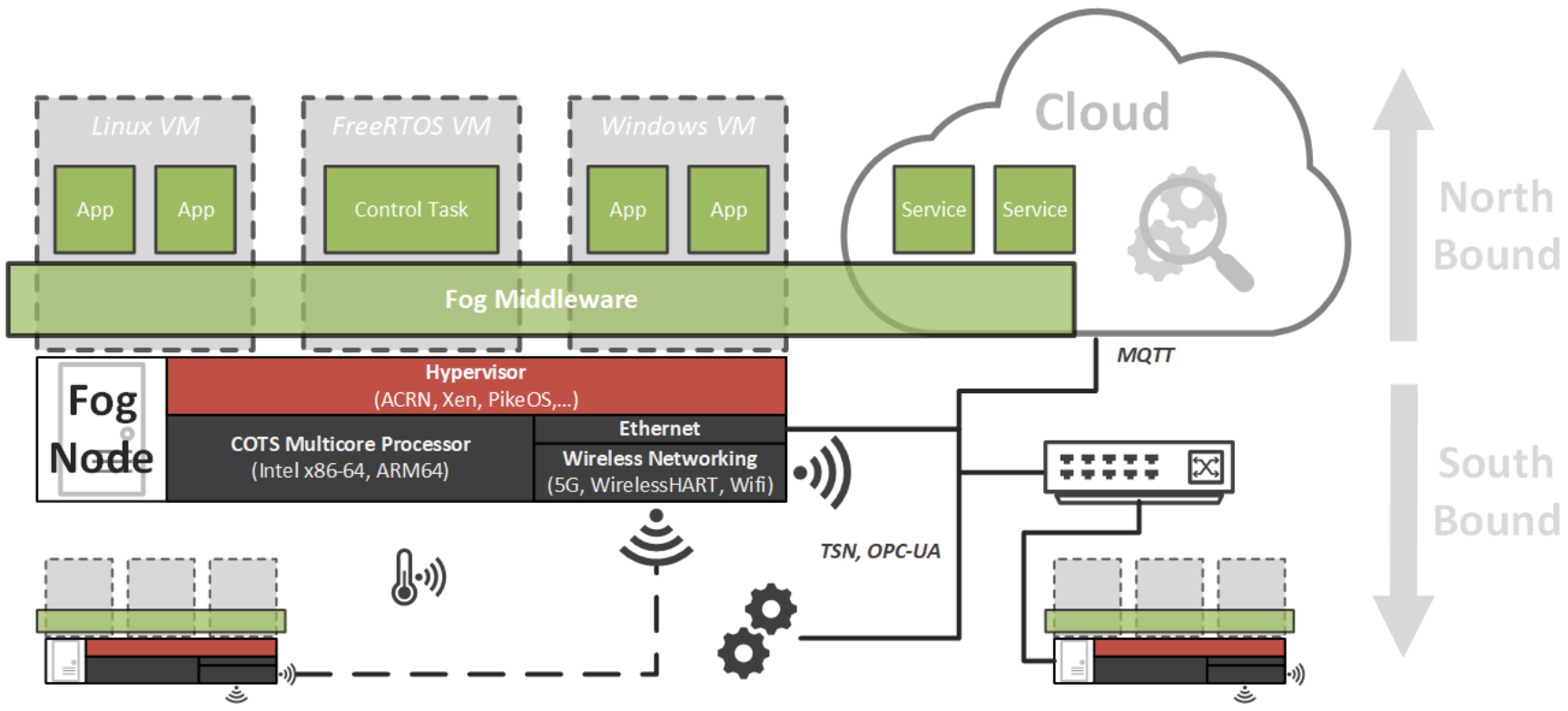

**Figure 4:** An illustration of the main components of the FORA Fog Computing Platform

We model the FORA FCP elements in a multi-layered approach, which consists of four layers, namely, Core, Software, Hardware, and Platform. The core layer represents the elements at a very abstract level. The Software and Hardware levels enrich the Core layer elements with the software and the hardware components. Finally, the Platform level composes and encapsulates the software and hardware components defined for each element at the Core layer into a single component that can be used to model a Fog Computing Platform.

## 4.1. Fog Node and communication

The FN envisioned in FORA comes with an Intel x86-64 or ARM64 COTS MCP that implements hardware virtualization extensions, such as Intel's VT-x and VT-d, Second Level Address Translation (SLAT), and Single-root I/O Virtualization (SR-IOV). Hardware virtualization extensions allow the hypervisor to host unmodified guest operating systems in VMs. Therefore, Intel VT-x (and its equivalent on ARM64) introduces an additional mode of operation with highest privileges to the instruction set architecture (ISA) that allows native execution of most sensitive instructions [26]. Intel's VT-d (and equivalent extensions on ARM) enable device passthrough that is granting a VM exclusive secure access to a PCI-e device. SR-IOV brings hardware support for sharing a physical PCI-e device with multiple VMs by offering so-called virtual functions. A chipset and PCI-e device with SR-IOV support allows sharing, e.g., a network card with multiple VMs, whilst guaranteeing isolation between the VMs. We do not require each FN having all of the just mentioned hardware extensions and features. The set of required features highly depends on its applications. Therefore, we differentiate three classes of FNs, which together make up the basis for the Fog Computing Platform. Note that the boundaries between the classes and their requirements are fluid:

- Class-1: FNs operating in very close proximity to machines and robots consisting of a multitude of sensors and actuators. The FN takes over critical hard real-time control tasks running in real-time VMs, hence it utilizes suitable COTS MCPs that come with less computational power yet a higher degree of determinism. Furthermore, the FN deploys a statically partitioned hypervisor, such as PikeOS [48], in order to fulfill strict timing requirements of its VMs. Southbound, the FN communicates via typical industrial field busses and/or TSN with OPC UA for non-critical communication. Northbound, thus in connection to other FNs, we leverage traditional Ethernet and TSN with machine-to-machine (M2M) protocols such as OPC UA and the ISO standard Message Queuing Telemetry Transport (MQTT).

- Class-2: FNs operating on the factory floor level. The FN does not take over critical control tasks, yet it can run soft real-time tasks, such as non-critical control, real-time data acquisition, data analysis, and data preprocessing. Therefore, the FN comes with a more powerful COTS MCP and a high degree of connectivity, including wired as well as wireless means of communication. The FN's hypervisor must be more dynamically configurable, such as ACRN or Xen, in order to be able to adapt to changes on the factory floor during runtime. However, soft real-time tasks might still be placed in statically configured VMs. An example for a class-2 FN is the Nerve MFN100 product from TTTech Computertechnik AG running the Xen hypervisor [90].

- Class-3: FNs operating on the factory or enterprise level. The FN collects operational data from all entities on the factory floors, either for sophisticated data analysis and process optimization, for long-term storage, or for forwarding data to the cloud. This requires a high degree of computational power and network throughput as provided by typical server processors and Gigabit Ethernet or even optical fiber since the FN effectively acts as cloud gateway. Southbound communication still involves TSN and OPC UA whereas northbound communicates solely bases upon TCP/IP and MQTT. The hypervisor must manage the FN's hardware resources dynamically and if need be even allow for over-provisioning of resources. Furthermore, integration with cloud services must be straightforward. Therefore, we use hypervisors that can usually be found in typical cloud environments, such as Xen or KVM.

Each FN utilizes hypervisor technology. A hypervisor is a low-level software layer that provides the abstraction of VMs to operating systems. A VM is a set of virtual resources such as, virtual CPUs, main memory, virtual I/O devices, or virtual time. The hypervisor manages the mapping of virtual to physical resources of all VMs it is hosting while guaranteeing strict isolation between its VMs. This allows a hypervisor to partition its hardware and run mixed critical applications isolated in dedicated VMs. Isolation does not come for free: State-of-the-art hypervisors for hard real-time applications, henceforth referred to as class-1 hypervisors, statically partition their resources, such as processor cores and time, main memory, and I/O devices, before runtime in order to achieve strict temporal and spatial isolation between VMs. As a result of their static configuration, class-1 hypervisors lack the flexibility to add, remove, and migrate VMs during runtime which make them less suited for more dynamic Industry 4.0 use cases. To that end we introduce a second class of hypervisors, namely class-2 hypervisors, that provide a good balance of temporal and spatial isolation and flexibility by utilizing mode changes [80] and compositional scheduling theory for the analysis of hierarchical scheduling [23], that is the scheduling of real-time tasks on virtual CPUs that in turn are being scheduled on physical cores of a MCP. Finally, there are FNs that take over high-level management tasks that require a high degree of flexibility. Therefore, they must be able to dynamically create VMs, migrate, and destroy VMs depending on the current task sets and their respective processing demand. We refer to these as class-3 hypervisors.

Furthermore, there are concepts that have to be considered for all three classes of hypervisors, such as the notion of a global time base that requires precise clock synchronization of FNs including VMs and hypervisors.

### *4.1.1. AADL models*

We have modelled the components discussed in the prvious section using AADL. The FCP is composed of a FN hardware platform and a virtualization solution, which can support both hypervisors and containers. The FN hardware platform has to be deployable in a wide range of industrial scenarios and support the execution of a variety of real-time system classes with different timing requirements [26]. An overview of the FN's design is presented in Fig. 5 and an illustration of the proposed FN's hardware architecture using AADL is presented in Fig. 6.

As mentioned, we utilize a hypervisor to guarantee temporal and spatial partitioning of the fog node hardware platform, see Fig. 5. The hypervisor provides virtual machines or partitions each of which runs their own operating system (OS). A privileged partition per fog node is in control of multiplexing access to shared devices by providing virtual devices. A virtual device consists of a frontend and a backend component whereas the frontend component, e.g., the virtualNetwork component, runs in an unprivileged partition and communicates with the backend component, e.g., the virtualNetworkBackend, in the privileged partition. We provide virtual devices for the I/O interface and the network interface. They connect via PCI to the fieldbus or the internal TSN switch of the FN hardware platform, as shown in Fig. 6.

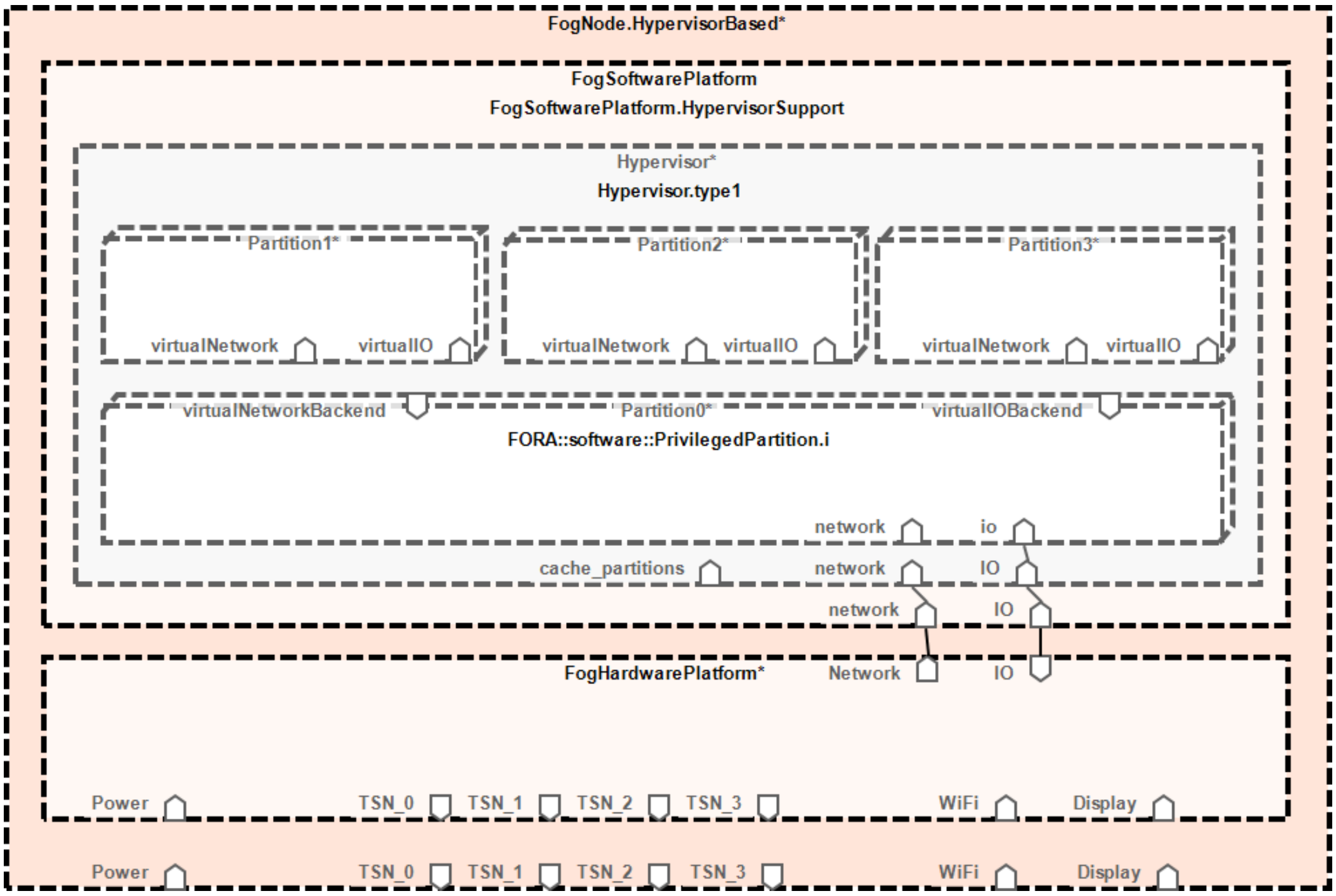

**Figure 5:** FORA Fog Node design overview

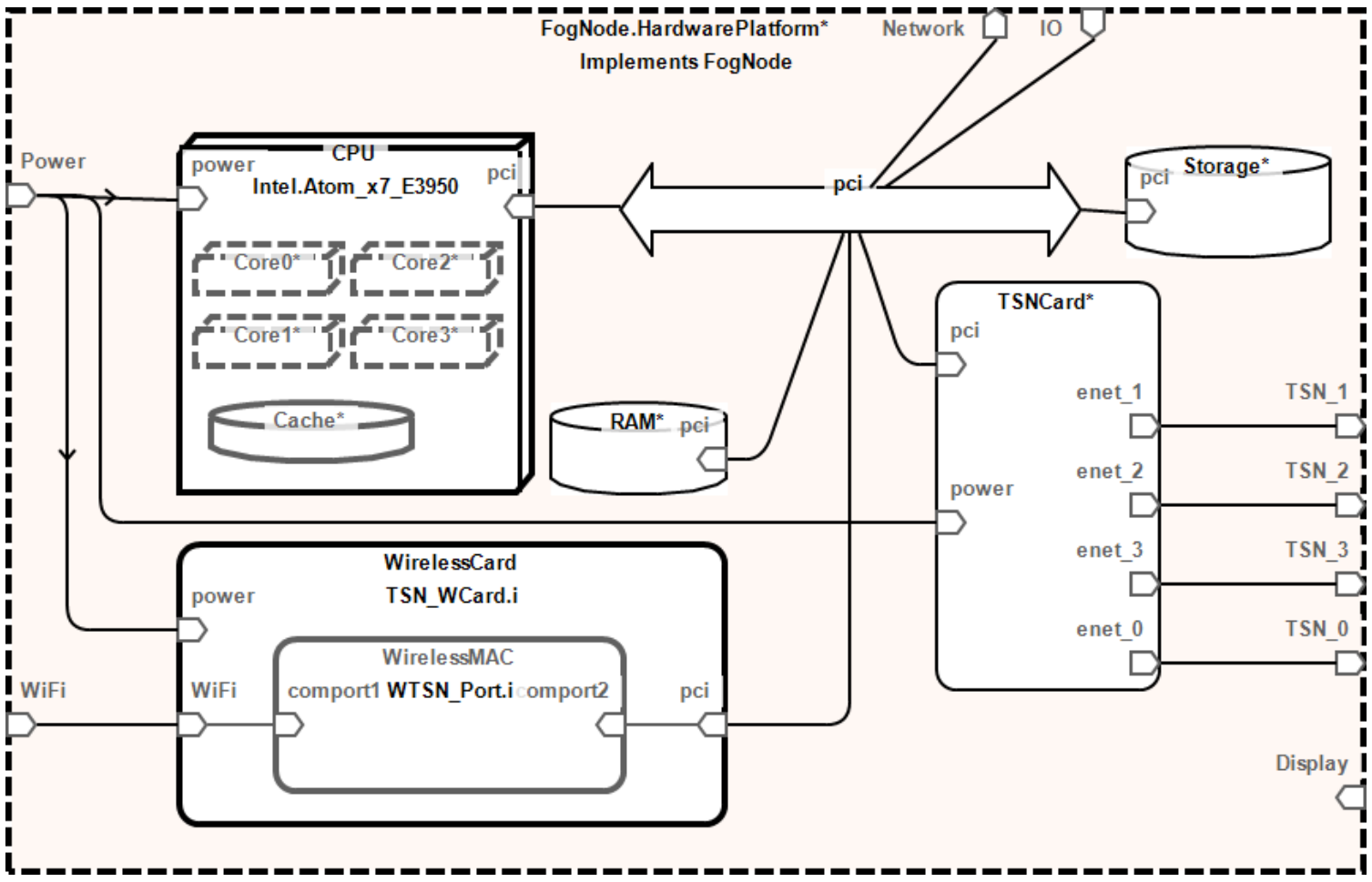

**Figure 6:** FORA Fog Node hardware architecture

The processor of the platform is envisioned to be an interchangeable component that can feature either a COTS MCP such as the Intel Atom shown in Fig. 6 or a more specialized real-time multi-core system able to provide bandwidth guaranteed core-to-core communication, time-predictability and static worst-case execution time analysis [80].

A fundamental aspect of Fog Computing is networking, that is why the proposed FN features three communication interfaces. For the communication paradigm, we propose the usage of TSN as it has been identified as the upcoming standard for real-time communication in Ethernet networks. It allows for mixed-criticality bounded latency communication, by separating traffic into priority classes. The arbitration of the traffic classes operates according to a schedule. To synchronize the schedule TSN employs a network-wide clock synchronization protocol, namely IEEE 802.1AS, that allows for sub-microsecond precision. Subsequently, the platform provides a TSN Ethernet switch device that enables the FN to communicate deterministically over Ethernet but also acts as an infrastructure node providing switching capabilities. A wireless TSN card allows for deterministic wireless communication. Finally, the platform provides a fieldbus interface, as a compatibility feature, to enable communication with common industrial devices and actuators. Special consideration is taken in the possible difference of the time domains between the TSN switch and the wireless TSN. The wireless TSN card is responsible for channel access for all wireless devices with a shared medium within a network under strict requirements on reliability and timeliness following different protocols such as contention-based protocols, contention-free protocols, and hybrid protocols [76]. Moreover, this component combined with a TSN card is to exchange data for the hybrid wired-wireless connections. Here, we consider a hybrid protocol TDMA-NOMA (Non Orthogonal Multiple Access) that can help to reduce end-to-end latency as well as to increase reliable communication.

## 4.2. Resource self-management, orchestration and self-configuration techniques

The FORA FCP reference architecture provides the means for resource self-management, orchestration, and self-configurations. Orchestration helps to align the resource demands of applications and the resource supply of different fog nodes with each other, i.e., to avoid that single applications utilize fog nodes in a greedy way [83]. Instead, resources are composed, i.e., combined with each other. It is the goal of the FORA FCP reference architecture to make sure that the overall system landscape is well-balanced and that all applications are provided with the necessary computational resources (via fog nodes). For this, it is taken into account which requirements the applications have, e.g., if they have real-time demands or not. Based on these demands, the orchestration and resource management capabilities of the fog nodes compute solutions for task scheduling and resource allocation. In general, orchestration and self-configuration techniques need to be provided on the system-level (i.e., for the fog nodes), and also take into account the networking perspective [49]. Importantly, resource allocation and task scheduling are not done on a single level, e.g., separately just at the edge of the network or just in the cloud. Instead, the FORA FCP provides cross-layer resource allocation, so that resources from the edge of the network to the cloud can be exploited if necessary and based on the demands of the applications.

As it is state-of-the-art in fog solutions, resource allocation is not done for fixed settings, but explicitly takes into account the volatility of fog landscapes, where nodes may enter or leave a landscape (or a network) at any point of time, and where the connections between nodes may also change during runtime [21]. For the resource allocation and task scheduling, functional and non-functional requirements are taken into account. The latter includes quality-of-service aspects like latency or security, but also the occurring cost of a particular resource allocation and task scheduling plan.

Resource allocation and task scheduling provide loosely-coupled functionalities, i.e., different methods and algorithms might be integrated into the FORA FCP, and it is even possible to provide transition mechanisms from one method/algorithm to another [4]. In contrast to other fog architectures, the FORA FCP explicitly foresees that fog landscapes may be organized in different ways, with hierarchical vs. fully decentralized landscapes being the two most extreme ways to organize a fog landscape [51].

While hierarchical landscapes are today the state of the art, fully decentralized landscapes more closely mirror the basic architecture of the IIoT. However, since most approaches to resource allocation and task scheduling are based on a hierarchical fog, novel approaches for decentralized landscapes need to be developed. To automate the distribution of applications and exploitation of computational resources in a fog landscape, the FORA FCP foresees that configuration tools are able to set up the single fog nodes based on the outcomes of the resource allocation and task scheduling computations, as well as further requirements.

Furthermore, for many of the functionalities mentioned here, it is necessary to monitor the nodes, in order to know their status. Hence, the FORA FCP provides the means to integrate monitors on different levels, e.g., for the cores or single tasks. Since fog (and IoT) landscapes are inherently volatile, faults may occur at any point in time. In order to be able to mask or mitigate failures in a fog landscape, the FORA FCP allows to allocate applications to new computational resources, even during the runtime of the system. For this, mechanisms which allow to store and re-establish the state of applications are necessary. Last but not least, fog landscapes should be able to be integrated with non-fog (legacy) systems. Especially in Industry 4.0 scenarios, OPC UA plays an important role [24], while the Data Distribution Service (DDS) is an important technology applied in real-time systems [68]. Therefore, the integration of fog nodes with OPC UA and DDS will be needed so that the FORA architecture does not provide a closed system, but is able to integrate other technologies to augment the fog, if meaningful.

#### 4.2.1. AADL models

The proposed fog platform has four major building blocks: (i) the hypervisor that can host OSes, virtual machines or containers, (ii) services for allocation and management of local resources and other essential services (e.g., TSN management), (iii) configuration services for the fog node, and (iv) the orchestration component that enables communication and resource sharing among fog nodes.

As mentioned, the platform is also capable of hosting *containers*, which are a lightweight virtualization alternative, e.g., Docker is a widely-used container technology for operating-system-level virtualisation [67]. Containers benefit from the fact that they share kernel functions of their host (i.e., they do not require separated operating systems running in each of the containers unlike virtual machines). This introduces the following advantages: (i) rapid boot time of containers, (ii) higher computational performance, and (iii) lower overhead. Containers offer self-healing mechanisms (i.e., prompt restarting of faulty containers) and mechanisms to increase dependability (containers can run in multiple instances). The main drawback of container-based virtualization is weaker resource isolation and potential lower level of security [59]. Although containers have typically been deployed on top of a rich OS, our FCP is also able to support them on top of separation kernels such as PikeOS; recent research has also extended containers with real-time capabilities.

In the following, we provide a detailed description of the components used in the AADL model from Fig. 7. The Control Configuration component provides the means for configuration which are used by the Node Management component to temporarily separate tasks with different criticality levels including the control tasks that have the highest criticality level, see [7] for more details. Additionally, the Node Management component contains a Security Management component. This component is responsible for the configuration of critical security mechanisms, such as the configuration of key distribution infrastructure. The component employs the system call instrumentation as proposed in [47] to monitor the entire system at runtime, which provides a configuration that determines the security functionality available to applications and services. It enables nodes to communicate with each other and external services securely, by defining protocols that can be used to uniquely identify peers, and to establish secure mutually authenticated communication channels. It also defines a set of acceptable data encryption algorithms for confidential data storage, in accordance with predefined security levels. These then are available as primitives to application services, enforcing node-wide consistent data security. Further, it manages the setup of firewalls, enabling host protection.

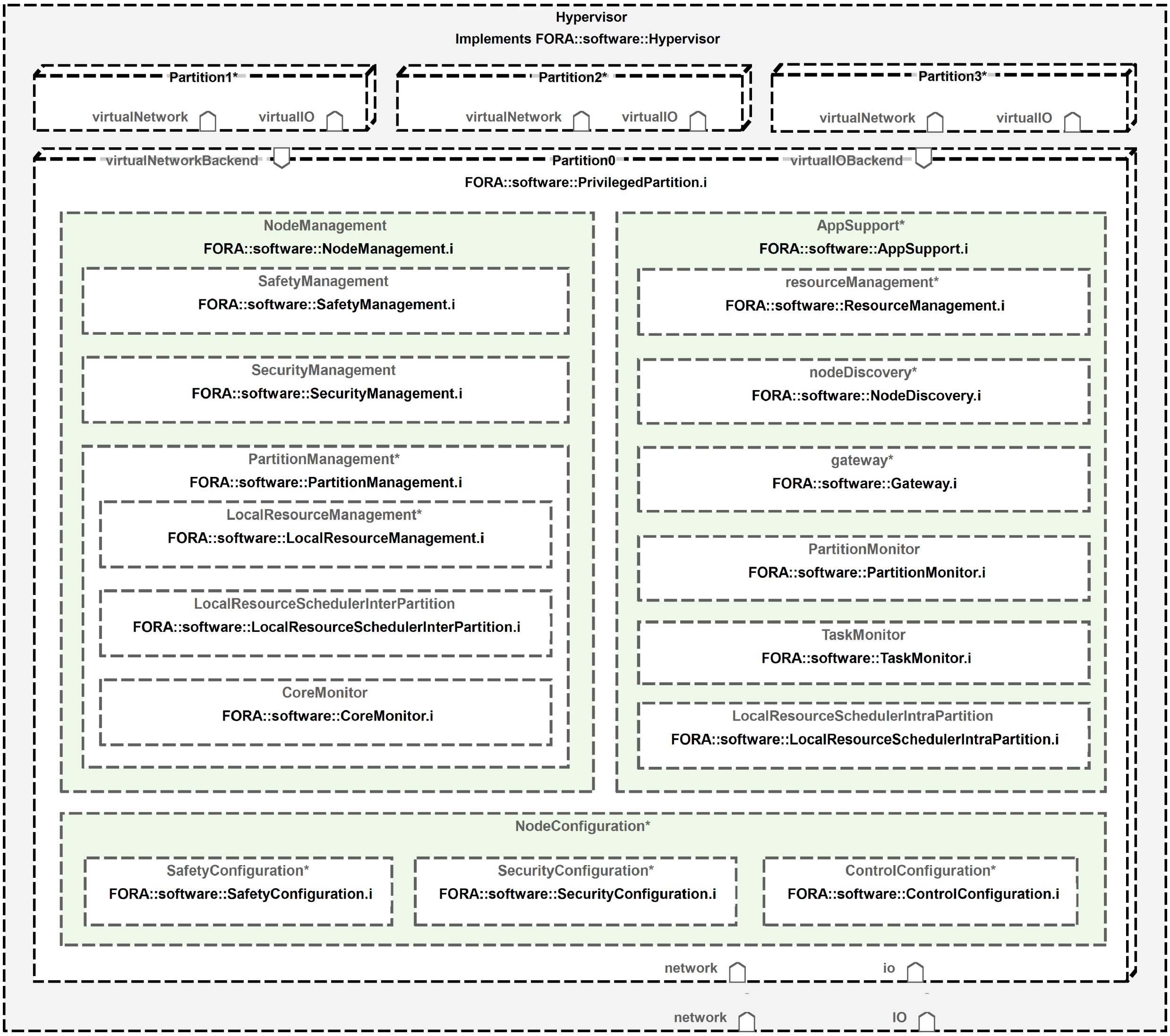

**Figure 7:** Overview of FCP's resource management, orchestration and configuration components

The AppSupport component provides the orchestration functionality to the FCP and consists of three different components, i.e., resource management, node discovery, and gateway. Each component contributes to the efficient use of the available resources in the network and the connection with different middlewares. A special focus lies in providing a deployment strategy for mapping the application's tasks to fog nodes, the discovery of the network topology, and establishing the communication between networks.

Resource management aims at deploying new applications on the FCP by creating a collaboration between fog nodes where each has the possibility of making local decisions regarding what tasks of the application to execute, see [6] for more details. This component consists of two distinct modules (i) a resource allocation module and (ii) a network monitoring module. The latter is set as a requirement for the FCP and aims at monitoring the network to provide the communication latency between different nodes. The former makes use of two new modules, i.e., the deployment module and the decision module, being responsible for finding a satisfiable mapping on the FCP for the deployed application. Besides the internal modules, the resource management component requires extra information about the fog nodes as well as providing access to different middlewares found in the network; information that is provided by collaborating with the remaining AppSupport components, i.e., node discovery and gateway.

Node discovery provides the resource management component with the candidate nodes for deploying applications. To do that, this component integrates node discovery algorithms such as [50], that store a set of neighbor nodes which can be used for distributing an application among various distributed fog nodes. Thus, node discovery is an essential part of the AppSupport component because it discovers all the fog nodes of the system, and makes their resources available to the resource management component which handles the deployment of the applications. Upon the discovery of new nodes, this component selectively chooses which nodes become neighbors. This is done based on proximity measurements in order to enable the execution of applications with low communication delay. Furthermore, the selection of neighbors determines the overall communication among the nodes, which can be either hierarchical or peer-to-peer. The hierarchical communication type enables the nodes to communicate in a tree-like topology which is based on layers. In the peer-to-peer communication type, the nodes communicate based on a flat model which does not use layers, see [52] for more details. The selection of the communication type is based on the requirements of the applications.

The gateway enables a fog node to communicate beyond its own network by connecting OPC UA and the DDS. After the resource management component deploys the application in a fog node, the user needs to configure the gateway by using a configuration file. The gateway configuration is static and is dependent on the configuration of the OPC UA and respectively DDS configuration. The OMG gateway specification document [40] provides further details about the gateway and its internal functioning.

The Core Monitor component monitors the core status. This component is a piece of software in charge of informing the Local Resource Management that a core is still alive. It is assumed that the core fail mode is not running. The implementation can be done in different ways, but in all of them, it must run from the privileged partition, as it must be aware of which partition is running and in which core. The core status information is essential to take local and global decisions in case of failure.

The Partition Monitor component monitors the partition status. This component has a similar function to the Core Monitor, but with the objective of monitoring the partitions. We assume here that the partition fail mode is a total failure, that means, the partition does not run anymore. Since this component must be aware of each existing partition and its status, it also must run from the privileged partition. This component can be part of the Core Monitor if the according monitoring objects are strictly connected, however, the information sent to the Local Resource Management has to be different The Task Monitor component monitors the critical task execution status and progress. Although the hypervisor can guarantee temporal and spatial task and partition isolation, in a COTS MCP, without previous detailed knowledge of all the tasks that can run at the same time in different cores, it is not possible to forecast how the inter-core interference delays the execution of the tasks due to the physical memory sharing among the cores. To this end, similar to the proposed approaches such as [12], we implement the Partition Monitor component for managing the shared resources on the platform. To guarantee critical task deadlines in this kind of hardware platform this component has to be implemented in such a way where it provides information to the Local Resource Management and Local Resource Scheduler Inter-Partition so that it can suspend non-critical partitions to avoid missing critical task deadlines.

The Local Resource Management component is in charge of gathering the information from the components described above, namely: Core Monitor, Partition Monitor, and Task Monitor. This component is also in charge of communicating the status of the resources to the following components: (i) Local Resource Scheduler Intra-Partition, (ii) Local Resource Scheduler Inter-Partition, and (iii) Control Configuration. This component must also run from the privileged partition since it has to have access to the other components in the entire platform.

Considering that the critical tasks and partitions are scheduled by the Control Configuration component, the Local Resource Scheduler Inter-Partition is in charge of finding feasible non-critical partitions scheduling according to the workload and locally available partitions. Taking into account the dynamic behavior of the task requests in a typical industrial automation scenario, this component is also in charge of controlling the non-critical partitions at runtime, suspending, running or changing the corresponding CPU occupation time. To schedule the partitions, this component has a strong interaction with the Local Resource Scheduler Intra-Partition.

Working tied to the Local Resource Scheduler Local Partition, the Local Resource Scheduler Intra-Partition component is responsible for finding a feasible non-critical task schedule according to the workload. As the previous component, to meet the dynamic requirements of a real industrial automation scenario, it is able to find a new schedule at runtime. This component is independent of the scheduling policy and it is possible to have different policies applied for different partitions.

## 4.3. Applications and services

The FCP has to be both agile and dependable. IT and OT worlds have different focus on dependability attributes such as safety and security, which have to be reconsidered in the converged IT/OT FCP. Regarding safety, when a system has the potential to harm humans or the environment (or is intended to mitigate or manage such harm), decision-makers require safety assurance evidence that it manages the risks acceptably. The conceptual basis for certification is that the evidence anticipates the possible circumstances that can arise from the interactions between the system and the environment, to show that these interactions do not pose an unacceptable risk. Hence, the FORA FCP integrates approaches, developed as platform services, for assuring the safety and security of the FCP. In addtion, the FN's proximity to the sensors and machines is an opportunity for improved data analytics, which, together with monitoring, improve agility when they support fast decision-making and resource allocation at the edge.

Security and privacy in OT lags behind IT, as the current state of practice is to use "air gaps", physically isolating sensitive equipment (locked doors and guards) from unsecured networks, which, with the introduction of standard IT systems no longer works. For IT, *security* and *privacy* are important, together with reliability, but safety is not considered. The convergence of IT and OT brings new security challenges, exposing previously isolated OT to new types of attacks [20]. Fog Computing introduces new security and privacy challenges that need to be addressed in order to promote such a new computing paradigm. Fog Computing inherits the Cloud Computing security issues, but these are more critical due to the safety issues of industrial systems. With the exception of very few preliminary papers, research is still immature [96].

*Big Data* and *data analytics* drive novel applications in Industry 4.0 [60]. Sensors and machines generate huge amounts of data: industrial data is growing faster than any other sector and manufacturing stores more data than any other sector. Access to industrial data is difficult because of the different data representations used by sensors and machines, data variety, and data velocity, e.g., the speed at which data is generated. Many applications enable data analytics by storing historical data in the Cloud, and running Machine Learning (ML) algorithms to gain insights that lead to value creation. Existing ML solutions from, e.g., IBM, Google, Microsoft, do not address industrial automation because large amounts of data cannot be moved to the Cloud due to bandwidth constraints, and with OT there are severe computation and storage limitations as we get closer to the machines. There is limited work in applying ML in real-time and on distributed data streams, e.g., distributed learning [2].

Regarding application development, the Cloud Computing programming model typically follows a Service-Oriented Architecture (SOA) model. Cloud applications are written using interacting (micro)services, developed as containers which rely on scalable Platform-as-a-Service (PaaS) [66]. Although there are many PaaS solutions for the Cloud, no such solutions exist for the Fog, where applications may reside in a continuum involving the machines, in the Fog Nodes, and the Cloud. In addition, critical applications with real-time properties cannot be modelled.

To address these challenges, our AADL platform model includes the following services, depicted in Fig. 8: resource monitoring services with extended insights into the individual nodes and network, and safety and security monitoring for recovering from failures and attacks; a new framework for edge analytics that goes beyond the traditional cloud-based post-collection analytics model that decouples data acquisition from knowledge extraction, improving thus the resource-efficiency of the FCP and IIoT end-devices, enabling real-time decision making, intelligent filtering, and resource prioritization; services for securing the FCP, based on the concept of Security-by-Contract and compliant with IETF MUD (Manufacturer Usage Description) [55], in order to be applied to the vast majority of IIoT devices, addressing the heterogeneity challenge.

### *4.3.1. AADL models*

The AADL components related to the discussion in the previous section are mainly of two categories, (1) Components that are essential for assuring the required levels of dependability attributes such as timeliness, safety and security and typically implemented in the platform, and (2) components that specifically target the application services and enable a coherent methodology for implementation of them. Category-1 involves some of the components that have already been explained as part of Sect. 4.1 and 4.2. We will contribute in designing as well as extending their scope w.r.t. enabling fault tolerance, timeliness, security and safety aspects. In this subsection, we focus on the components that are of Category-2, i.e., implementing application-level services in Fig. 8.

We are developing appropriate intrusion detection techniques to secure the execution of mixed-criticality applications on the fog node. The intrusion detection module collects system events, such as hardware performance counters and system calls [47]. It analyzes them with Machine Learning techniques to detect anomalous patterns of execution. The solution is non-intrusive, since it observes the monitored software's interface without direct interaction and hence will not affect the predictability in any adverse manner. The intrusion detection performance overhead is limited due to the embedded environment properties. Because it is deployed locally on the target, its execution should not interfere with other system applications, which potentially have safety requirements and deadlines. It also must not be reachable by other untrusted entities to preserve its own security. Therefore, the secure integration of intrusion detection module in a partitioning based system design for embedded mixed criticality environments is also essential. These techniques are implemented partly in the hypervisor to have direct hardware access (see Fig. 7) and partly in the Security service in Fig. 8.

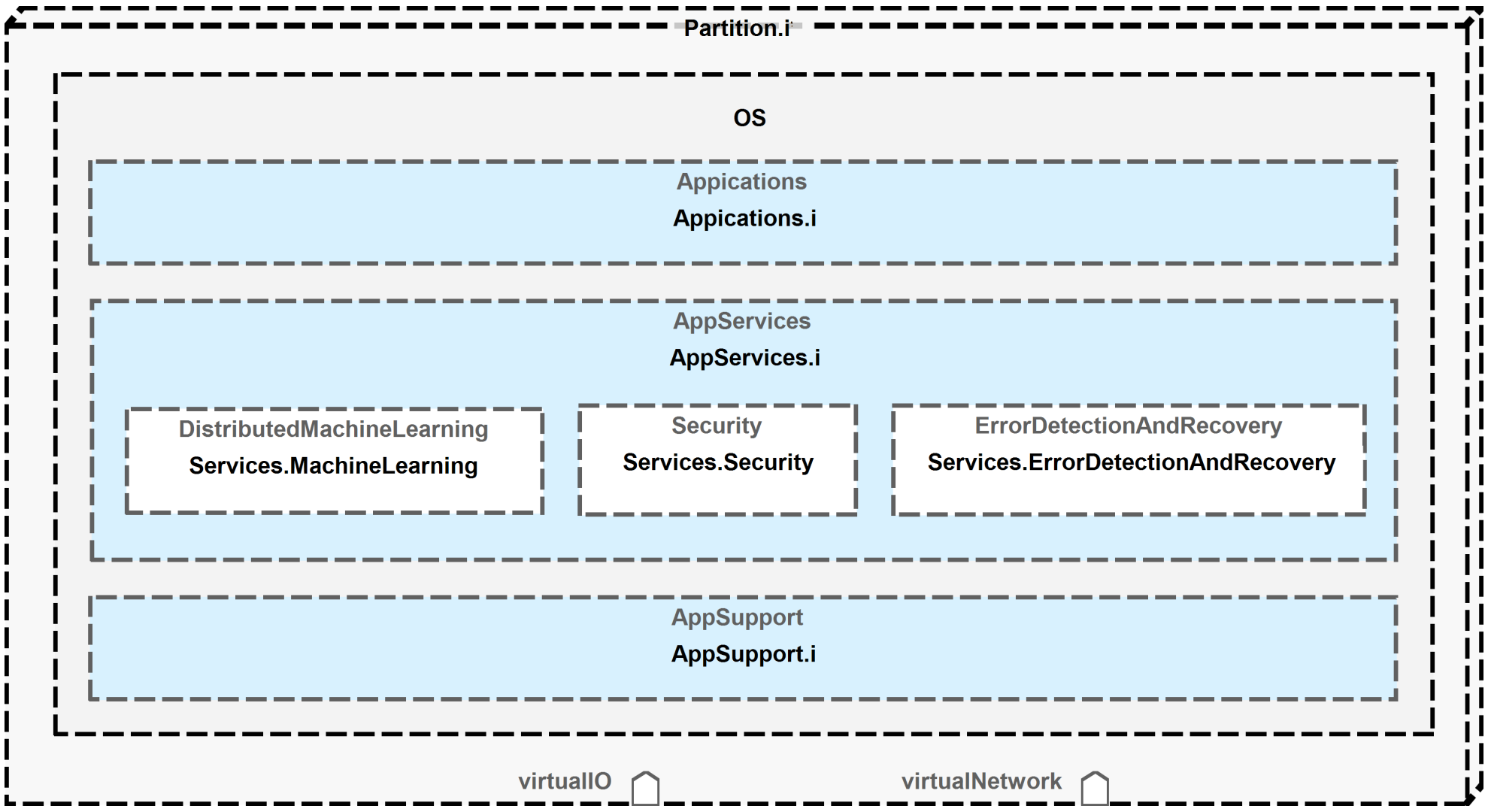

**Figure 8:** Overview the FCP's services

Due to the increasing focus on TSN in the provision of predictable network traffic, we are developing a fault detection, isolation, and recovery (FDIR) method to be applied in the context of TSN communication and task's executions, similar to [29]. The method will reside in the middleware and will monitor tasks executions and network communication traffic. It applies fault detection and identification techniques, and enables appropriate recovery mechanisms in case of faults. IEEE 802.1CB TSN standard provides fault-tolerance by means of stream redundancy i.e., splitting streams across disjunctive links to maximize probability of correct reception in the presence of link faults. We are also exploring the feasibility of a Fault Tolerant Communication Configurator, which is capable of configuring the network stream transmissions in an optimal manner to ensure that resources are conserved while at the same time, fault-tolerant guarantees are provided. Similar to the security techniques, these fault-tolerance techniques are implemented both at lower levels (within the hypervisor and RTOSes) and in the ErrorDetectionAndRecovery service depicted in Fig. 8.

A fog platform brings the computing power from the remote cloud-side closer to the edge devices to reduce latency, as the unprecedented generation of data causes ineligible latency to process the data in a centralized fashion in the Cloud. In this new setting, edge devices with distributed computing capability, such as sensors and surveillance cameras can communicate with fog nodes with less latency. Furthermore, local computing (at edge side) may improve privacy and trust. Hence, we integrate a new method to decompose the data processing, by dividing them between edge devices and fog nodes, intelligently. We apply active learning on edge devices; and federated learning on the fog node which significantly reduces the data samples to train the model as well as the communication cost, similar to [73]. This work has been used to implement a Predictive Maintenance framework, which uses distributed machine learning, where the distributed drivers and the centralized server jointly (collaboratively) train one global model. Typically, the decentralized drivers placed in different locations, generating data that captures the local information instead of global information and they train their local models based on partial knowledge. The aggregation step at the server-side enables the information sharing between drivers and server to obtain one model with overall knowledge. Then, the server sends the aggregated model back to drivers. The edge analytics service is implemented in the Machine Learning component from Fig. 8.

Regarding mixed-criticality application modeling, critical and non-critical applications can be defined in FORA

FCP architecture as AADL process and thread components. Non-critical applications can be defined as standard process and thread components in AADL, whereas critical applications should be defined as extensions of CriticalApp and CriticalTask components introduced by FORA AADL models, which are extensions of process and thread components. Critical applications may be control applications with quality-of-control requirements, safety-critical applications with dependability requirements (e.g., have to be replicated for fault-tolerance reasons) or real-time applications with soft or hard real-time properties. All of these applications are real-time and are modelled using the typical sporadic task model from real-time theory [17], where each periodic task has a period (minimum inter-arrival time for sporadic tasks), a worst-case execution time and a deadline. We opted to use the timing properties proposed by SAE in the Timing_Properties property set, which have been developed for capturing such timing requirements. Fig. 9 shows an example of modeling a critical application with three critical tasks. The AADL source code for this application with the detailed specifications for one of its tasks is presented in Listing 1 to show how the timing properties of an application and a task can be specified.

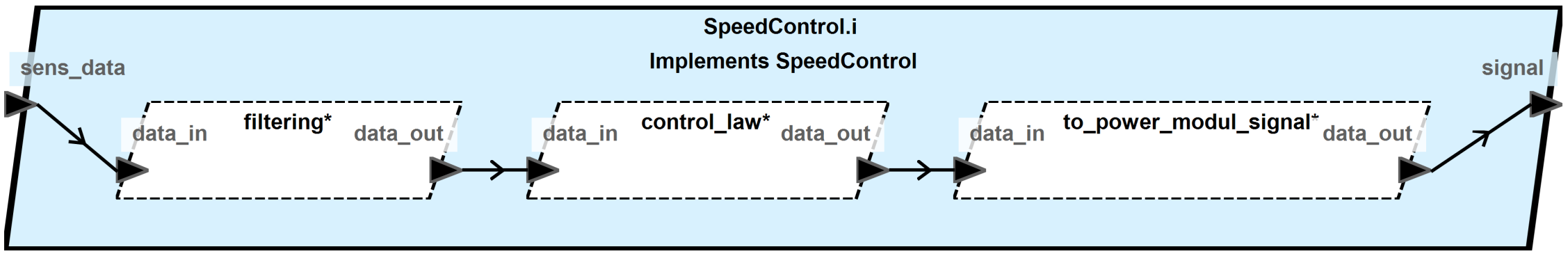

**Figure 9:** An example of modeling a critical application with three critical tasks using FORA AADL models

```
process SpeedControl extends FORA::core::CriticalApp
features
  sens_data : in data port sens_data {arinc653::sampling_refresh_period => 10 ms;};
  signal : out data port signal {arinc653::sampling_refresh_period => 10 ms;};
properties
    FORA::core::CriticalApp::SIL => 3
end SpeedControl;

process implementation SpeedControl.i
subcomponents
  filtering : thread filtering;
  ...
connections
  c0 : port sens_data -> filtering.data_in;
  c1 : port filtering.data_out -> control_law.data_in;
    ...
end SpeedControl.i;

thread filtering extends FORA::core::CriticalTask
features
  data_in : in data port sens_data;
  data_out : out data port filter_data;
properties
   dispatch_protocol => periodic;
   period            => 10ms;
   deadline          => 10ms;
   compute_execution_time => 100ms..200ms;
end filtering;
```

Listing 1: Part of an AADL source code to specify a critical application with three critical tasks

## 5. Evaluation

Sect. 3 has outlined the methodology used for the definition and evaluation of the proposed Fog Computing Platform reference architecture. In this section we report the results obtained when modeling, implementing and evaluating a *Conveyor Distribution System* Use Case (UC1 mentioned in Sect. 3). The details of UC1 have been presented in [8]. Here, we extend that work to: show how the AADL from Sect. 4 can be used to model the UC; show how the FORA Technology Bricks can be plugged into this AADL model to implement a prototype; evaluate the ability of successfully implementing the UC using the proposed reference architecture. Use Cases UC2 and UC3 mentioned in Sect. 3 have been reported in [78, 28], but without a focus on AADL modeling and evaluation.

In UC1, a conveyor distribution system is used to distribute packages from an inventory to different destinations. The conveyor distribution machine is well-known and widely used in inventories for the automatic distribution of packages. The machine is fed with packages from one side and reads the tag of the received package. It gets the destination of the package by accessing a database with the read tag and drives the package towards the destination from one of the other sides of the machine.

Conveyor distribution uses electric motors and drives. *Electric drives* alter the frequency and voltage of an electric motor's current for different rotation speed, torque, and position of its shaft using the implemented real-time software controlling the power electronic circuits [13]. Industrial controllers (e.g., PLCs) sitting on the "Control level" of the automation pyramid determine the required output of electric motors which sit on the "Machine level" (see Fig. 1). The electric drives placed close to electric motors on the "machine level" convert the determined output to a corresponding electric current to drive the electric motors. Thus electric drives produce massive and critical data about the controlling machinery. Sending all the data to other computing nodes over the Internet or the control network would consume bandwidth and it is also discouraged by the factory owners for confidentiality concerns. Hence, in this use case, we extend the electric drives to serve as fog nodes, leading to new offerings like programmability, local analytics, and connectivity to customer Clouds.

We model the UC's architecture with AADL by using the FORA AADL components from Sect. 4 and refining them to reflect the UC requirements. We model positioning and tag-scanner sensors as *FORA::Platform::Sensor*, and the electric motors as *FORA::Platform::Actuator*. We model the TSN switches as *FORA::Platform::TSN_Switch* and configure them using FORA AADL property set developed for configuring a TSN network, which allows specifying routing specifications and message schedules for a switch as system properties. We model the fog nodes by extending *FORA::Platform::FogNode* and configure them by refining the software and hardware subcomponents of the *FogNode* model. Thus, Fig. 10 presents the AADL model of a Fog Node. The FN consists of a hardware platform which has a multicore processor, a TSN enabled network switch and I/O to an external power module for generating the electric current; and a software stack which has a hypervisor, a middleware, and partitions with dedicated operating systems (real-time OS for running control applications and Linux/Windows-based OS for running best effort applications) and application layers, respectively. The fog nodes are assumed to run mixed-criticality applications including control applications for controlling electric motors. The mixed-criticality applications are temporaly isolated using partitions which are managed by the hypervisor i.e., PikeOS in this UC which supports static partition tables (see [7] for more information about partition tables and enforced isolation). Finally, we model the UC platform as a system that consists of the required instances of the aforementioned components connected via TSN. Fig. 11 presents the resulted AADL architecture model for the UC. As it can be seen, the architecture consists of three switches ($SW_i$, denoted in the following text with the notation $\mathcal{W}$), five fog nodes ($FN_i$, denoted with $\mathcal{E}$), five position sensors and one tag scanner sensor (all denoted with $\mathcal{S}$). Each fog node is connected to an electric motor ($m_i$) it controls.

We model the applications using AADL as discussed in Sect. 4.3.1. For space reasons, instead of showing the full AADL model, we instead use a table where we present the properties of the applications, see Table 2. In this UC each fog node $\mathcal{E}_i$ runs a control application which receives a message from its corresponding sensor $S_i$ and controls the belt to drive the package by determining the required corresponding electric motor's output. The fog node also runs an electric drive application which sets the electric motors current for the external power module via I/O. Thus, the UC has five control applications (one for each fog node), and each control application gets one stream as input and sends its output 10 ms after receiving the input (we assume that the outputs have fixed offset from the inputs and have no latency).

The details of the applications are shown in Table 2. The applications 1 to 5 are control applications for controlling the speed of electric motors, and hence have control performance requirements. Each application has a criticality Level (L)—that can represent the Safety Integrity Levels (SIL) of the application, with values from 0, non-critical, to 4, highest criticality [87], a number of tasks with the same period (P), and a computation cost (C) which is the sum of worst-case execution times of the tasks over their period. We assume that the all the applications (including the control applications) interact via a set of streams which have hard real-time requirements, and all the streams are prioritized concerning their criticality. The details of the streams (size $S$, period $T$ and routing) are shown in Table 2. We assume that all links have the speed of 100 Mbps. Listing 1 presents the AADL specifications for one of these applications (speed control).

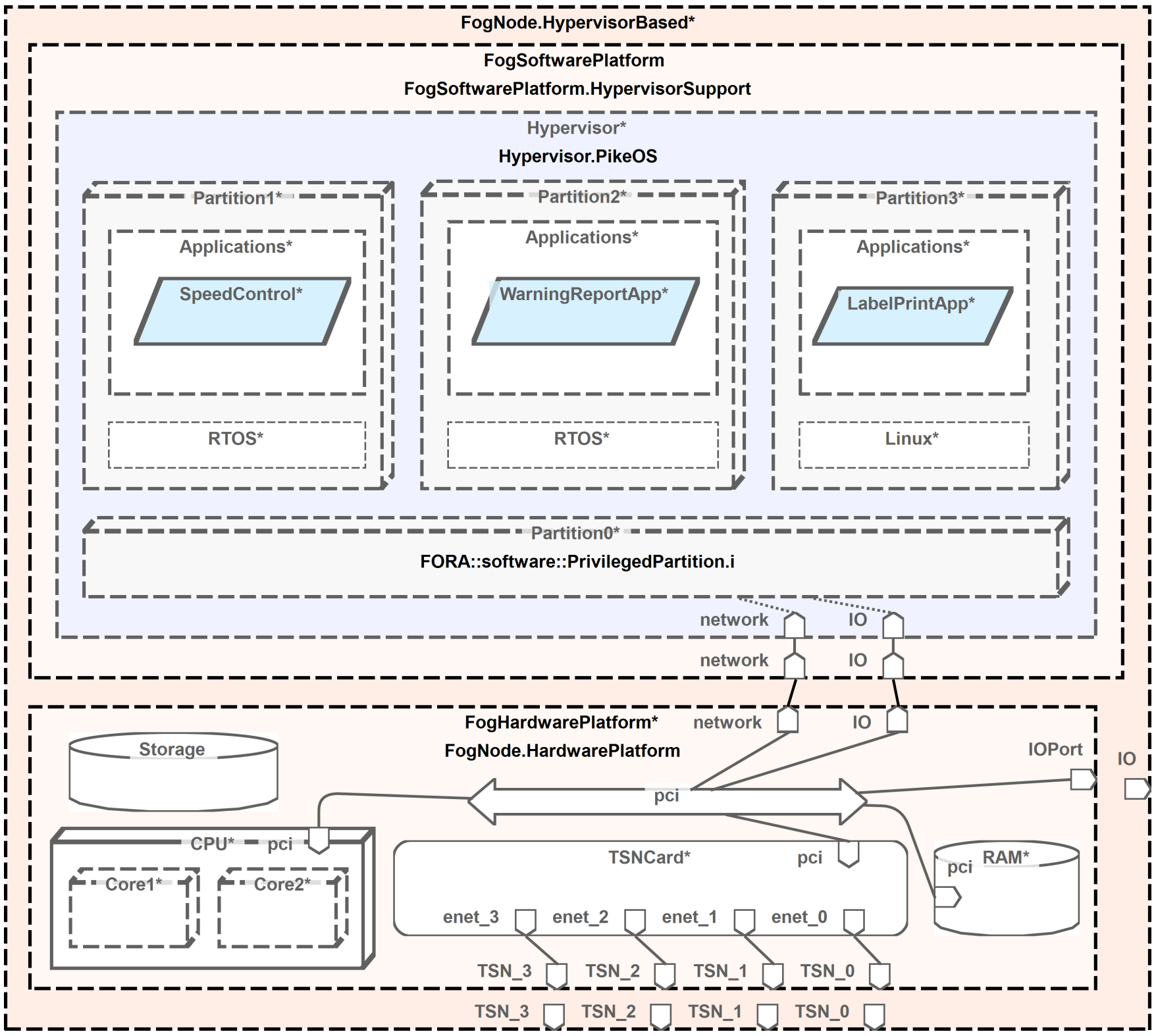

**Figure 10:** Configuration of a fog node in the UC's architecture

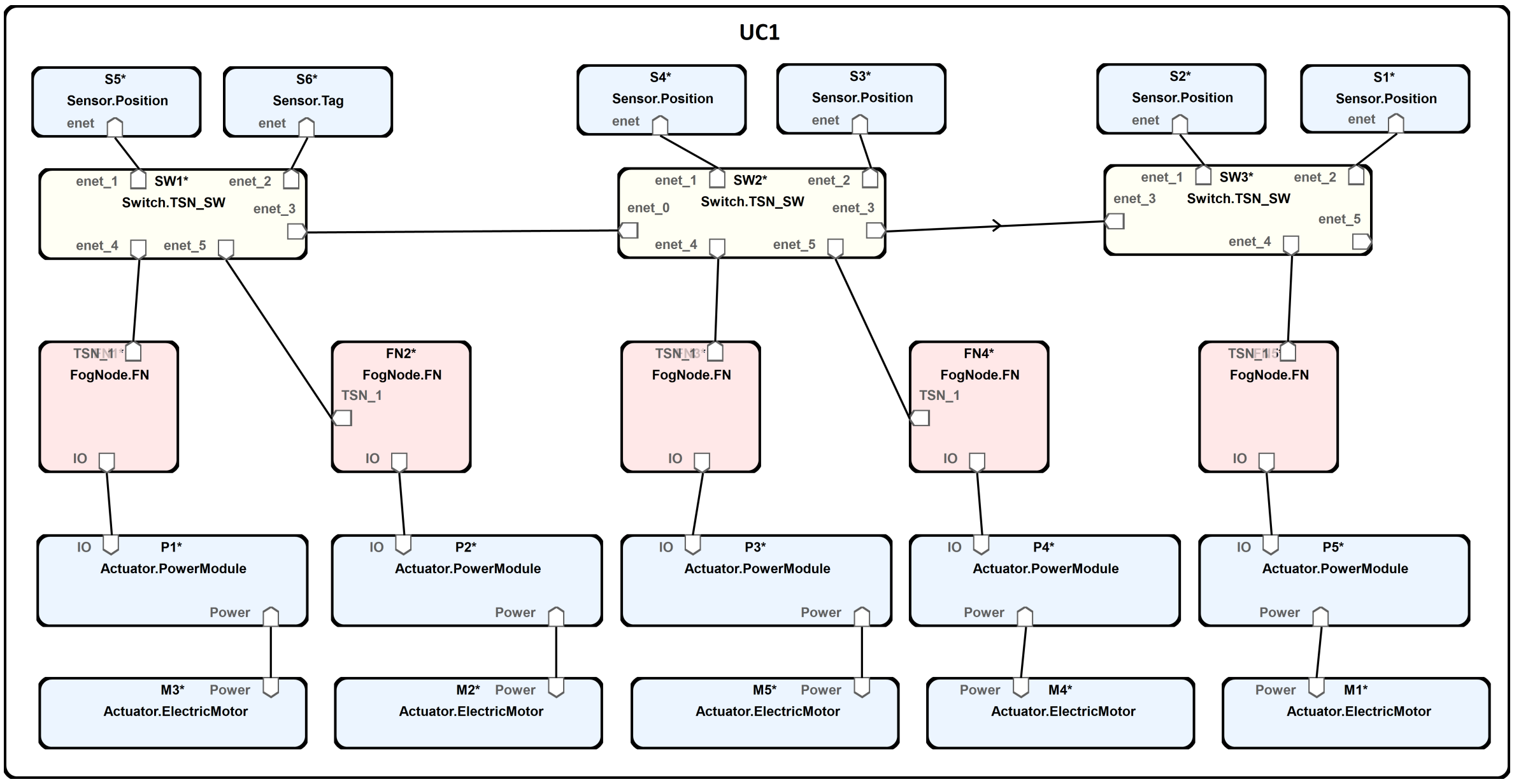

**Figure 11:** AADL diagram of the UC's architecture

**Table 2**
UC1: applications, streams and evaluation results

| Application | L | tasks | P (ms) | C | relevant stream | S (bytes) | T (ms) | routing | ED ($\mu$s) | ED after TESLA ($\mu$s) |
|---|---|---|---|---|---|---|---|---|---|---|
| $m_1$ control | 3 | 3 | 10 | 0.35 | $\mathcal{S}_1$ data | 700 | 10 | $\mathcal{S}_1 \rightarrow \mathcal{W}_1 \rightarrow \mathcal{E}_1$ | 60 | 1241 |
| $m_2$ control | 3 | 3 | 10 | 0.35 | $\mathcal{S}_2$ data | 850 | 10 | $\mathcal{S}_2 \rightarrow \mathcal{W}_1 \rightarrow \mathcal{E}_2$ | 72 | 1481 |
| $m_3$ control | 3 | 3 | 10 | 0.35 | $\mathcal{S}_3$ data | 600 | 10 | $\mathcal{S}_3 \rightarrow \mathcal{W}_2 \rightarrow \mathcal{E}_3$ | 52 | 1111 |
| $m_4$ control | 3 | 3 | 10 | 0.35 | $\mathcal{S}_4$ data | 950 | 10 | $\mathcal{S}_4 \rightarrow \mathcal{W}_2 \rightarrow \mathcal{E}_4$ | 80 | 2407 |
| $m_5$ control | 3 | 3 | 10 | 0.35 | $\mathcal{S}_5$ data | 500 | 10 | $\mathcal{S}_5 \rightarrow \mathcal{W}_3 \rightarrow \mathcal{E}_5$ | 44 | 921 |
| Package status | 2 | 2 | 10 | 0.3 | $m_2$ state | 1100 | 20 | $\mathcal{E}_2 \rightarrow \mathcal{W}_1 \rightarrow \mathcal{E}_1$ | 152 | 1911 |
| Motor break | 3 | 3 | 10 | 0.35 | – | – | – | – | – | – |
| Table break | 2 | 1 | 8 | 0.31 | – | – | – | – | – | – |
| SCADA | 1 | 4 | 10 | 0.28 | $\mathcal{E}_5$ data | 920 | 10 | $\mathcal{E}_5 \rightarrow \mathcal{W}_3 \rightarrow \mathcal{W}_2 \rightarrow \mathcal{E}_4$ | 254 | 2389 |
| User interface | 1 | 3 | 6 | 0.28 | – | – | – | – | – | – |
| Database access | 1 | 8 | 15 | 0.59 | $\mathcal{S}_6$ data | 1200 | 30 | $\mathcal{S}_6 \rightarrow \mathcal{W}_3 \rightarrow \mathcal{W}_2 \rightarrow \mathcal{E}_3$ | 200 | 3091 |
| Weight report | 2 | 5 | 15 | 0.47 | – | – | – | – | – | – |
| Warning | 2 | 2 | 10 | 0.26 | $\mathcal{E}_4$ data | 700 | 50 | $\mathcal{E}_4 \rightarrow \mathcal{W}_2 \rightarrow \mathcal{E}_3$ | 260 | 1751 |
| Destination set | 1 | 3 | 8 | 0.56 | $m_2$ set | 850 | 50 | $\mathcal{E}_3 \rightarrow \mathcal{W}_2 \rightarrow \mathcal{W}_1 \rightarrow \mathcal{E}_2$ | 144 | 2221 |
| Label print | 1 | 4 | 12 | 0.59 | – | – | – | – | – | – |

## 5.1. Implementation model of network traffic for QoC

The UC's architecture is configured to guarantee the timing requirements of all the network streams on TSN, including the control applications' streams which have more stringent timing requirements. The configuration is composed of Gate Control Lists (GCLs) for the TSN network switches which represents the message schedules, and is provided by the Control Configuration component (see Sect. 4.2.1). The GCLs regulate the network traffic such that functional timing requirements of network streams i.e., stream deadlines, and their non-functional timing requirements are met. The control performance requirements of control applications is defined as Quality-of-Control (QoC) (see [10] for more details). We evaluate these requirements by analyzing the GCLs.

We employ a Constraint Programming-based schedule synthesis strategy, implemented as an OSATE plug-in, aiming at maximizing the QoC and satisfying the deadlines of network streams, proposed in [10] to generate the GCLs. Thus, all the streams have been successfully scheduled, i.e., none of the deadlines is missed. The configuration has also provided the minimum delay and jitter for the streams, resulting in a good control performance. The column 10 (last but one) in Table 2 shows the maximum end-to-end delay (ED) of streams. We used JitterTime [19] to simulate the behavior of the control applications which reports an average value of 0.009 for the QoC of all control applications, i.e., a good control performance (see [10] for the exact cost function).

## 5.2. Implementation model of hypervisor partitions and task schedules for QoC

Mixed-criticality applications sharing the same platform have to be isolated from each other, otherwise a faulty lower-criticality task may interfere with a higher-criticality task, leading to failure. In the UC's architecture, each fog node uses a deterministic hypervisor for virtualizing the applications by providing deterministic access to shared resources via a static configuration table. The deterministic access realizes the temporal isolation for the applications with different levels of criticality via partitioning aiming at protecting the applications from the possible interference.

The Control Configuration component (see Sect. 4.2.1) uses the heuristic algorithm proposed in [10] to allocate a partition for each criticality level of the assigned applications, and to decide mapping of the partitions to cores of the fog nodes. The component generates schedules for partitions and for tasks inside each partition considering the determined mapping of tasks to the partitions and cores. The generated task schedules are optimized for QoC of control applications which are assumed to have high-criticality levels.

We evaluate the performance of the configuration in providing temporal separation and preserving QoC for control applications. We assume that each fog node has a dual-core processor. The Control Configuration component has determined the mapping of partitions to the cores of fog nodes. The component has also successfully scheduled the partitions and all the tasks inside the allocated partitions. The results show that none of the tasks has missed its deadline and all of the tasks are isolated concerning their criticality levels by mapping them to the partitions with same criticality levels. The results show that the cores have average utilization value of 57.2% and the maximum utilization value is reported as 73.6% for the core 0 of the fog node $\mathcal{E}_3$. Furthermore, the control application has a good control performance, which is evaluated with JitterTime [19] that calculates a value of 0.4103 for the QoC (cf. the cost function from [7]).

### 5.3. Addressing extensibility for dynamic fog applications

The UC consists of statically allocated critical applications to run on the platform. These applications are statically mapped to the cores and partitions, and scheduled inside the mapped partitions at design time. The UC's architecture is also capable of running dynamic non-critical applications which can migrate in-and-out of the fog nodes over time and be removed/replaced by other such applications. This capability, known as extensibility, is realized in a way that the design time configuration is not modified which is a necessity for keeping the performance level of the statically allocated critical applications as well as avoiding the safety re-certification of critical applications [9].

The extended configuration is provided by the Node Configuration component (see Sect. 4.2.1) at runtime and schedules the dynamic non-critical applications on their arrival. To allow more dynamic non-critical applications to be added at runtime without negatively impacting the performance of existing applications, the design time configuration needs to be optimized for extensibility which is realized by distributing the idle time of the design time schedules. The idle time spaces of these schedules are used to accommodate tasks of dynamic applications. Less-deviated idle time duration enables the fog nodes to accommodate more dynamic non-critical applications.

The design time extensible schedules are generated using the method proposed in [9] and implemented as an OSATE plug-in. We evaluate the extensibility of the UC's schedules by optimizing the generated schedules in Sect. 5.2. We take the schedule $s_8$ (representing the schedule on the second core of fog node $\mathcal{E}_4$) and depict it in Fig. 12a (as "BASE") using a Gantt chart, where the boxes are execution slices, and the arrows show task preemption. The execution slices are denoted with the task's number. All the tasks have the same criticality level and are scheduled inside the same partition. While "BASE" is not optimized for extensibility, we give an example for optimized version of the same schedule described in Fig. 12c as "OPTIMIZED". We use the same extensibility metric as presented in [9], which reports the values for deviation of idle time duration as 0.0014 and 0.0003 for "BASE" and "OPTIMIZED"

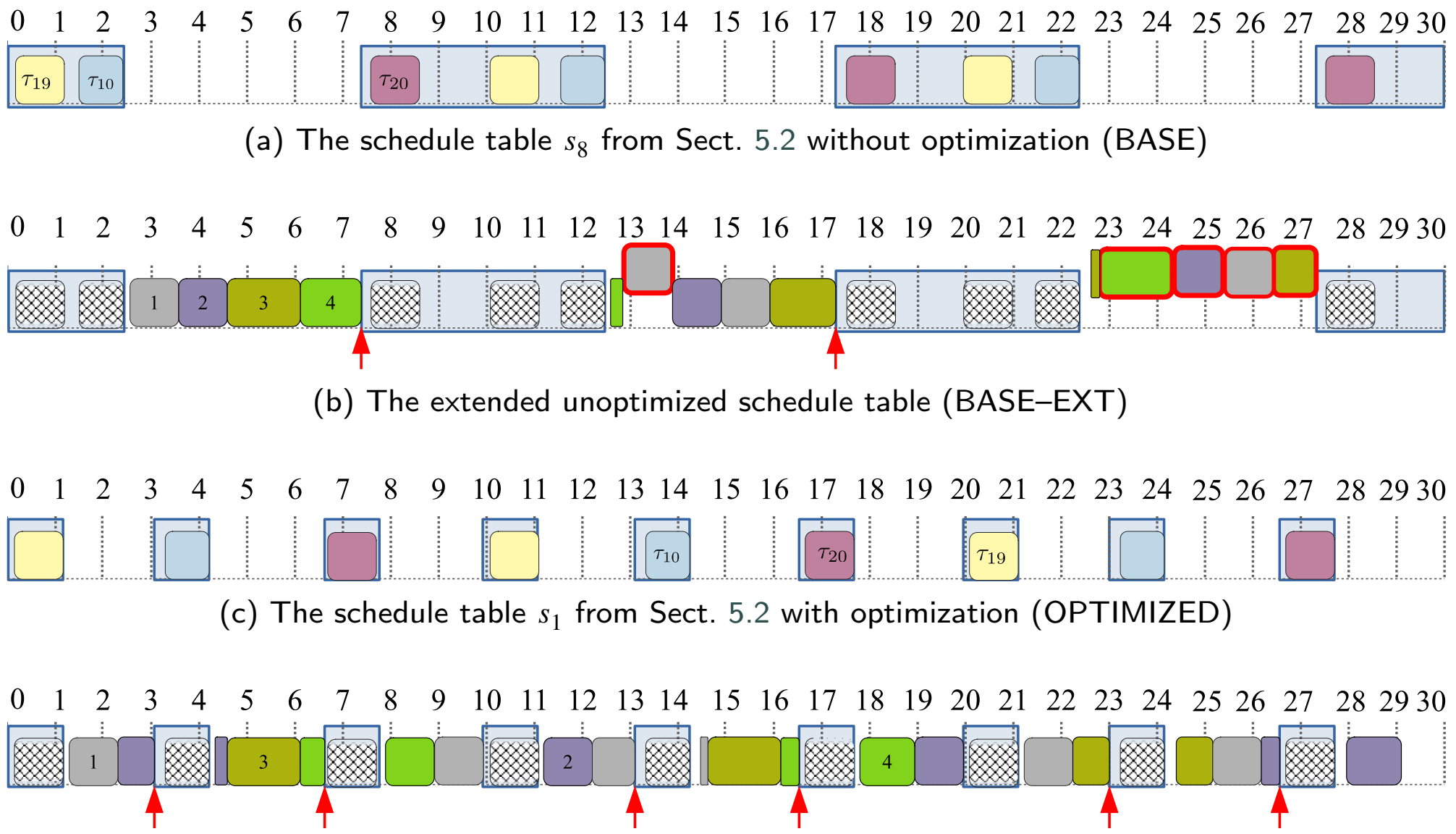

(a) The schedule table $s_8$ from Sect. 5.2 without optimization (BASE)

(b) The extended unoptimized schedule table (BASE–EXT)

(c) The schedule table $s_1$ from Sect. 5.2 with optimization (OPTIMIZED)

(d) The extended optimized schedule table (OPTIMIZED–EXT)

**Figure 12:** Four schedule tables for extensibility example: The colored boxes are execution slices; the transparent boxes show the partitions; the red bordered boxes show missed-deadline tasks. The arrows show preemption and the hatched boxes shows occupied time slots.

respectively, showing that "OPTIMIZED" has less-deviated idle time duration, i.e., it is more extensible.

We consider a scenario where engineers want to add four dynamic non-critical applications to the fog node $\mathcal{E}_4$. These applications represent logging applications. Each dynamic application composed of a single task which has a deadline constraint equal to its period. The applications' periods are 6, 8, 10, 12 ms and their computation costs are 17%, 13%, 15%, 13%, respectively.

Thus, we extend the schedules "BASE" and "OPTIMIZED" by adding the dynamic applications 1–4 to the schedules. The resulting schedules are depicted in Fig.12b and Fig.12d as "BASE-EXT" and "OPTIMIZED-EXT" respectively. The results show that the dynamic application has successfully scheduled in "OPTIMIZED-EXT" i.e., none of the applications has missed its deadline, whereas some deadlines are missed in "BASE-EXT", for example, the task of the application 1 has missed its deadline at 12,900 ms. The configuration provided by Node Management component shows promising results in successful adding of dynamic applications without modifying the existing schedules, and is able to bring extensibility to the schedules of fog nodes.

### 5.4. Addressing security requirements in TSN

The UC's architecture requires secure communication for confidential messages to guarantee confidentiality, integrity of the data, and authenticity of the remote party. In the UC, the platform blocks any attempt at communication to endpoints it cannot authenticate.

Since TSN does not provide security mechanisms, we employ the Timed Efficient Stream Loss-Tolerant Authentication (TESLA) [77] to guarantee the security requirements in our proposed architecture. TESLA is a low-resource multicast authentication protocol which relies on synchronized schedules of tasks and messages which is implemented in the Security Configuration component of the AADL model. The authentication protocol uses MD5-MACs of 16 Byte Length and 16 Byte keys. The configuration provided by the component consists of a dedicated security application composed of two tasks, for each network message which is scheduled to run either at the message's reception or transmission depending on the end system that the application is running on.

We scheduled the UC's streams from Sect. 5.1 using the TELSA method to evaluate the security mechanism of the UC's architecture. The results are shown in Table 2. The Columns 10 and 11 (last two) exhibit the maximum end-to-end delay (ED) of streams before and after adding the security mechanism respectively. The security configuration provided confidential authenticated communication by using MD5-MACs and authentication keys at the expense of increasing the end-to-end delay of messages by 1723 $\mu$s on average.

## 6. Conclusions and discussion

This paper has presented a Fog Computing Platform (FCP) reference architecture aimed at Industrial IoT applications. The architecture was defined and evaluated within an overall methodology that was driven by requirements collected via three IIoT use cases. The definition of the FCP reference architecture has been done via AADL models. We have presented an overview of the models and the entities of the FCP within three main themes: (i) computing device and communication, (ii) resource management and orchestration and (iii) application and services. We have discussed the reasoning and analysis behind the definition of the FCP and listed the major components and the technology bricks developed to implement a design.

The proposed reference architecture was evaluated on a conveyor belt distribution system demonstrator, showing the capability to successfully model and implement IIoT applications. As future work, the FORA AADL models will be further refined and aligned to standards (e.g., IEEE 1934 OpenFog).

A Fog Computing Platform brings several benefits to Industrial IoT applications: End-users benefit from machine interoperability and resource elasticity. The FCP scales on demand to meet business needs and connects all assets of end-users to enable data capture. End-users will be able to connect the machines, the Fog Nodes and the Cloud, allowing optimal resource allocation, driving costs down and value up. Dependable middleware and interoperability protocols make data available for innovative applications, e.g., data-driven decision-making, data analytics.

The FORA FCP provides services needed to rapidly develop, securely deploy and efficiently operate industrial applications. The software platform provides standardization across multiple silos and enables businesses to quickly take advantage of operational and business innovations.

The FCP infrastructure meets stringent industrial regulatory requirements (safety and security), which cannot be met by public Clouds. This reduces the security risks with networked machines. The new virtualized FCP handles security incidents such that the critical operations are not impacted, reducing downtime. New security services help end-users deploy secure industrial applications and detect abnormal or suspicious behavior, recovering from attacks and reducing losses. Data analytics services offer insights, enabling decision-making to increase asset utilization, deploy servicing and maintenance resources efficiently to lower repair costs, plan performance improvements.